\documentclass[]{interact}

\usepackage{epstopdf}
\usepackage[caption=false]{subfig}	
\usepackage{color}

\usepackage[numbers,sort&compress]{natbib}
\bibpunct[, ]{[}{]}{,}{n}{,}{,}
\renewcommand\bibfont{\fontsize{10}{12}\selectfont}
\makeatletter
\def\NAT@def@citea{\def\@citea{\NAT@separator}}
\makeatother

\theoremstyle{plain}

\theoremstyle{definition}

\theoremstyle{remark}

\usepackage{gensymb}
\newcommand{\abs}[1]{\lvert#1\rvert}

\begin{document}

\articletype{ARTICLE TEMPLATE}

\title{{Nonlinear dynamics in breathing-soliton lasers}}

\author{
\name{Junsong Peng\textsuperscript{a,b,c}\thanks{CONTACT J. Peng. Email: jspeng@lps.ecnu.edu.cn},  Xiuqi Wu\textsuperscript{a,c}, Huiyu Kang\textsuperscript{a}, Anran Zhou\textsuperscript{a}, Ying Zhang\textsuperscript{a}, Heping Zeng\textsuperscript{a,b,c,d}, Christophe Finot\textsuperscript{e}\thanks{CONTACT C. Finot. Email: christophe.finot@u-bourgogne.fr} and Sonia Boscolo\textsuperscript{f,g,h}\thanks{CONTACT S. Boscolo. Email: s.a.boscolo@aston.ac.uk}}
\affil{\textsuperscript{a}State Key Laboratory of Precision Spectroscopy and Hainan Institute, East China
Normal University, Shanghai 200062, China; 
\textsuperscript{b}Collaborative Innovation Center of Extreme Optics, Shanxi University, Taiyuan, Shanxi 030006, China; 
\textsuperscript{c}Chongqing Key Laboratory of Precision
Optics, Chongqing Institute of East China Normal University, Chongqing 401120, China; 
\textsuperscript{d}Chongqing Institute for Brain and Intelligence, Guangyang Bay Laboratory, Chongqing 400064, China;
\textsuperscript{e}Universit\'e Bourgogne Europe, CNRS, Laboratoire Interdisciplinaire Carnot de Bourgogne ICB UMR 6303, F-21000 Dijon, France; \textsuperscript{f}Aston Institute of Photonic Technologies, Aston University, Birmingham B4 7ET, UK; \textsuperscript{g}VPIphotonics, 10587 Berlin, Germany; \textsuperscript{h}Istituto Nazionale di Fisica Nucleare, Laboratori Nazionali del Sud (INFN-LNS), 95125 Catania, Italy}
}

\maketitle

\begin{abstract}
We review recent advances in the study of nonlinear dynamics in mode-locked fibre lasers operating in the breathing (pulsating) soliton regime. Leveraging advanced diagnostics and control strategies---including genetic algorithms---we uncover a rich spectrum of dynamical behaviours, including frequency-locked breathers, fractal Farey hierarchies, Arnold tongues with anomalous features, and breather molecular complexes. We also identify a novel route to chaos via modulated subharmonic states. These findings underscore the utility of fibre lasers as model systems for exploring complex dissipative dynamics, offering new opportunities for ultrafast laser control and fundamental studies in nonlinear science.
\end{abstract}


\section{Introduction}
\label{Intro}
Mode-locked fibre lasers are valued not only as compact sources of ultrashort pulses but also as highly controllable laboratories for nonlinear science. In these cavities, a delicate balance among dispersion, Kerr nonlinearity, and wavelength-dependent gain and loss sustains {\itshape dissipative solitons}---localised structures that persist through continuous energy exchange with their environment \cite{NP_Grelu_2012, OC_Grelu_2024, OLT_Singh_2025}. Because key parameters such as pump power, intra-cavity dispersion, and saturable-absorber characteristics can be tuned with precision, fibre lasers provide an archetypal platform for exploring dissipative dynamics. Their accessible parameter space supports a rich palette of behaviours: breathing oscillations \cite{PRL_Sotocrespo_2000,PRE_Chang_2015,SA_Peng_2019}, soliton explosions \cite{PRL_Sotocrespo_2000, PRL_Cundiff_2002,Optica_Runge_2015,CP_Peng_2019,OL_Liu_2016}, chaotic \cite{OL_Horowitz_1997,PRL_SotoCrespo_2005} and rogue-wave states \cite{PRL_Lecaplain_2012,LPR_Wu_2023}, harmonic mode locking, and self-organised patterns ranging from soliton bunches \cite{OL_Grelu_2002} to stable multi-soliton bound states (``soliton molecules'') \cite{PRA_Malomed_1991,PRL_Stratmann_2005,Science_Herink_2017,PRL_Krupa_2017,PRL_Liu_2018,LPR_Peng_2018,NC_Wang_2019,Optica_Liu_2022,APN_Zou_2025}.
Many of these phenomena were identified decades ago, yet their underlying physics is only now being elucidated thanks to advanced single-shot diagnostics \cite{NP_Goda_2013, NP_Mahjoubfar_2017, RPP_Wang_2020,APX_Godin} that resolve pulse evolution on a round-trip basis. Insights from these experiments advance fundamental theory---by supplying a testbed for far-from-equilibrium models---and guide practical design, informing the optimisation of next-generation ultrafast sources. This dual relevance places mode-locked fibre lasers at the centre of contemporary research in ultrafast photonics and nonlinear dissipative systems.

Breathing (or pulsating) solitons, manifesting as localised temporal or spatial structures exhibiting periodic oscillations in energy, are fundamental nonlinear modes observed across a wide range of physical systems. They appear in various domains of natural science, including condensed matter physics \cite{PRA_Su_2015}, fluid dynamics \cite{PRX_Chabchoub_2012,PRL_Chabchoub_2016}, plasma physics \cite{PRL_Muller_1999}, chemistry, molecular biology, and nonlinear optics \cite{OE_Dudley_2009,SR_Kibler_2012,PRX_Frisquet_2013}.
In conservative systems, breathing solitons can arise spontaneously via modulation instability of continuous waves \cite{PRA_Su_2015,SR_Toenger_2015,OL_Hammani_2011,PRL_Xu_2019}, owing to the fact that the governing evolution equations---such as the nonlinear Schr\"odinger equation (NLSE), sine-Gordon equation, or Korteweg–de Vries  equation---admit periodic or quasi-periodic solutions \cite{Akhmediev_1997,RP_Copie_2020}. In contrast, in dissipative systems such as passive or active optical cavities, breathing solitons typically emerge as limit cycles originating from a steady state through a Hopf bifurcation, also referred to as a Poincar\'e–Andronov–Hopf bifurcation \cite{Strogatz_2015}, when system parameters are varied. In optics, dissipative breathers---initially studied experimentally in passive Kerr fibre cavities \cite{OE_Leo_2013} and microresonators \cite{NC_Yu_2017,NC_Lucas_2017,PRL_Cole_2019}---have also emerged as a universal mode-locking regime in ultrafast fibre lasers \cite{SA_Peng_2019,OL_Du_2018, PRL_Xian_2020,PRApp_Peng_2019}. In particular, in \cite{SA_Peng_2019}, we reported the first real-time experimental observation of single breathers and breather-pair molecules in a laser cavity using advanced real-time detection techniques. Since then, a series of distinct experimental studies on breather structures in laser systems have been published by various groups (e.g., \cite{OE_Wang_2019, OE_Chen_2019, OE_Luo_2020, LPR_Liu_2020, OL_Wang_2020, LPR_Peng_2021, PRL_Guo_2021, LPR_Wu_2022, LPR_Krupa_2022, OL_Du_2022, APR_Zhou_2022, PRL_Wu_2023,LPR_Wang_2023,LP_Lu_2023,PRL_Cui_2023,PRL_Kang_2024, OLT_Fan_2024, OLT_Li_2024, LPR_Wang_2025,SA_Wu_2025}).

This sustained interest is mainly due to two key factors. First, breathing solitons represent a novel mode-locking regime in lasers. Their understanding, characterisation, and optimisation may open new frontiers in ultrafast laser physics. In particular, self-synchronisation phenomena observed in lasers supporting breathing solitons \cite{NC_Wu_2022, SA_Wu_2025, PRL_Wu_2023} provide critical insights into ultrafast laser dynamics---knowledge that is essential for the development and practical deployment of next-generation laser systems.

Second, breathing-soliton lasers offer an excellent platform for uncovering novel nonlinear dynamics in dissipative systems. In linear oscillator ensembles, synchronisation manifests primarily as phase-locking between coupled modes and is fully described by linear superposition, with each mode evolving independently. In breathing-soliton lasers, synchronisation emerges intrinsically from nonlinear coupling among the cavity’s internal frequencies, enabling self-synchronisation without external forcing. These nonlinear interactions also allow chaotic attractors, which are prohibited in finite-dimensional linear systems, where dynamics are confined to linear combinations of non-interacting eigenmodes and no mechanism exists for intrinsic frequency locking. In this paper, we provide a review of key findings from our recent research in this rapidly advancing field, situating breathing-soliton dynamics within the broader context of nonlinear science. Our aim is not to present an exhaustive survey, but to emphasise conceptual and mechanistic understanding of the breathing-soliton phenomenon, the associated research methods, and the latest advancements, and to delineate how these results advance the understanding of complex dissipative dynamics. These include: the emergence of higher-order Farey hierarchies of frequency-locked breather states and self-similar fractal dynamics \cite{NC_Wu_2022}; the appearance of abnormal synchronisation domains (unusual Arnold tongues) \cite{SA_Wu_2025}; transitions between synchronised and desynchronised breather regimes, including the identification of a novel intermediate dynamic state \cite{PRL_Wu_2023}; and a new route to chaos through the breakdown of regular dynamics \cite{PRL_Kang_2024}. We also demonstrate the use of genetic algorithms (GAs) to generate breather dynamics with controlled characteristics \cite{LPR_Wu_2022,NC_Wu_2022}.

The paper is organised as follows:
Section \ref{sec:Methods} reviews the key experimental and theoretical tools and methodologies that enable advanced studies of breathing solitons in ultrafast fibre lasers, with an emphasis on real-time diagnostics and genetic-algorithm control.
Section \ref{sec:Synchronisation} examines synchronisation dynamics, including frequency-locked breathers, Farey hierarchies, and Arnold-tongue structures. 
Section \ref{sec:Complexes} extends this discussion to multi-breather complexes and demonstrates their intelligent control.
Section \ref{sec:Chaos} explores transitions of breathing-soliton lasers from regular to chaotic dynamics.
Finally, the concluding section summarises our findings and outlines future research directions.

\section{From measurement to model: Laser architecture, diagnostics and control strategy}
\label{sec:Methods}

In this section, we review the key tools and methodologies that have enabled the advanced investigation of breathing solitons in ultrafast fibre lasers. Addressing this problem required overcoming several challenges. On the experimental side, substantial progress was achieved through the development of reliable and controllable laser platforms. A critical component was the implementation of appropriate detection techniques, as the non-stationary nature of breathers necessitates diagnostics capable of resolving pulse evolution on a round-trip basis. These measurement capabilities, in turn, facilitated feedback mechanisms for optimising cavity parameters. To interpret and predict breather dynamics, it was equally important to compare experimental observations with numerical simulations. For this purpose, we examine two principal modelling frameworks.

It is important to note that this discussion is limited to directly observable breathing solitons. Vectorial structures whose periodic behaviour only becomes apparent after polarisation analysis \cite{OL_Du_2020, OE_Li_2020, OLT_Wang_2023, APN_Huang_2023} are not addressed here, as their study necessitates polarisation-resolved models. We also exclude regimes associated with dissipative soliton explosions, characterised by abrupt yet quasi-periodic variations in pulse properties linked to Q-switching dynamics \cite{Optica_Runge_2015, OL_Liu_2016, PRApp_Peng_2019}.

\subsection{Experimental setup}
Here, we summarise the key experimental techniques employed for the generation, observation, and control of breathing solitons. Detailed descriptions of setups and methodologies can be found in our previous publications \cite{SA_Peng_2019, LPR_Peng_2021, LPR_Wu_2022, LPR_Wu_2023, NC_Wu_2022, PRL_Kang_2024, PRL_Wu_2023}.

\subsubsection{Fibre laser architecture}
\label{sec:laser_setup}

\begin{figure}
\centering
{
\resizebox*{13.cm}{!}
{\includegraphics{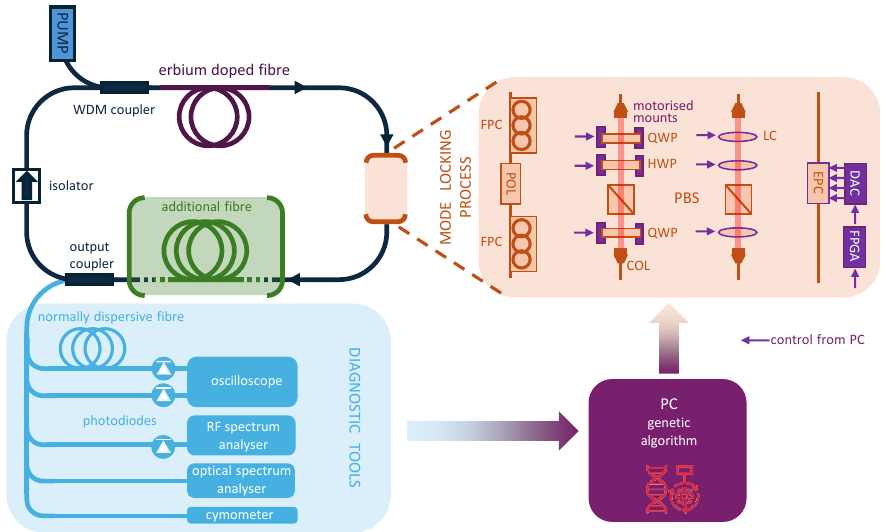}}}\hspace{5pt}
\caption{\textbf{Experimental setup}. Schematic of a typical fibre laser cavity used to generate and characterise breathing solitons. The setup includes a set of diagnostic tools for detailed observation of the pulsating structures, as well as passive and active components enabling mode locking via nonlinear polarisation rotation. FPC, fibre polarisation controller; POL, polariser; COL, collimator; QWP/HWP, quarter-/half-wave plates; PBS, polarisation beam splitter; LC, liquid crystal phase retarder; EPC, electronically driven polarisation controller; DAC, digital-to-analog converter.} \label{fig:Experimental_setup}
\end{figure}

The generation of breathing solitons in fibre lasers does not inherently require cavity designs that differ radically from those used for the generation of conventional ultrashort dissipative solitons. Figure~\ref{fig:Experimental_setup} illustrates a typical ring-cavity architecture employed to generate such structures at telecommunication wavelengths (around 1550$\,$nm, within the C band). All components operate in the single-mode regime and are widely available commercially.
While the discussion here focuses on a unidirectional ring cavity, breathing solitons have also been observed in figure-eight lasers \cite{OL_Deng_2021, PRL_Kang_2024}, figure-nine lasers \cite{OLT_Yuan_2025}, linear cavities \cite{SR_Akosman_2018}, and Mamyshev oscillators \cite{PRA_Cao_2022}, highlighting the universality of the breathing regime beyond a specific laser configuration.

In the ring-cavity architecture, gain around 1550$\,$nm is provided by an erbium-doped fibre, pumped by a continuous-wave laser diode operating at 976$\,$nm. Although most reported results have been achieved in the C band, oscillatory dynamics have also been demonstrated at other wavelengths, including 1.03$\,\mu\mathrm{m}$, 1.6$\,\mu\mathrm{m}$, 1.7$\,\mu\mathrm{m}$, and 2$\,\mu\mathrm{m}$, using different doped fibres---such as ytterbium-doped \cite{OL_Suzuki_2018, OL_Deng_2021, LPR_Krupa_2022, OE_Qi_2023, LP_Fu_2024}, erbium-doped  \cite{OL_Wang_2020, OLT_Li_2022}, thulium-holmium co-doped \cite{Chaos_Ji_2024}, and thulium-doped fibres \cite{PJ_Wang_2017, Chaos_Liu_2023}.
An inline isolator is typically employed to enforce unidirectional operation, although breathing dynamics have also been reported under bidirectional operation \cite{OL_Zhou_22}. Polarisation control is achieved by incorporating a fibre polariser or a polarisation beam splitter, which ensures maintenance of a single polarisation state within the cavity. These components may also be combined in the form of a polarisation-dependent isolator.
The cavity elements---such as output couplers, pump wavelength-division multiplexers, and isolators---are generally based on standard single-mode fibre, which exhibits anomalous dispersion (typically around $-22.8\,\mathrm{ps^2/km}$ at 1550$\,$nm). The net cavity dispersion can be finely tuned by inserting segments of normal-dispersion fibre. Breathing dynamics have been observed and characterised in both net-anomalous and net-normal dispersion regimes \cite{PRE_SotoCrespo_2004, OL_Du_2018}, each exhibiting distinct temporal and spectral features. The oscillation period of breathing solitons varies strongly with cavity parameters and pump power---ranging from a few to several tens of cavity round-trips near zero dispersion, to periodicities on the order of hundreds of round-trips in cavities operating under normal dispersion conditions \cite{SA_Peng_2019}.

A fundamental aspect of any ultrashort pulse laser is the mechanism that enables mode locking, particularly the choice and implementation of the saturable absorber within the cavity \cite{JSTQE_Haus_2000}. This absorber may be a physical component that exploits the intensity-dependent absorption properties of a material \cite{OL_Wang_2018, PRA_He_2021, OLT_Yuan_2024}. However, to circumvent issues such as material degradation and provide greater control and flexibility, {\itshape virtual} saturable absorbers can also be employed. In such cases, the Kerr nonlinearity of silica---manifesting through effects like nonlinear polarisation rotation (NPR) or phase modulation---is converted into effective intensity modulation by means of polarisation filtering or auxiliary optical loops.
Ultrafast lasers utilising NPR \cite{EL_Matsas_1992} have achieved notable success and remain widely used due to their simplicity and high performance. Nevertheless, tuning operating regimes in NPR-based systems has historically relied on manual adjustment of fibre-based polarisation controllers or combinations of discrete waveplates. This empirical approach limits reproducibility and stability.
To address these limitations, a range of externally controllable polarisation management techniques have been developed. These include waveplates mounted on motorised stages \cite{SR_Lapre_2023}; conventional mechanical three-loop fibre polarisation controllers enhanced with electronic control \cite{OE_Zheng_2021}; and electronically driven polarisation controllers composed of three or four fibre squeezers oriented at $45\degree$ with respect to each other \cite{APB_Hellwig_2010, OL_Shen_2012, LPR_Wu_2022, LPR_Wu_2023}. Another approach involves the use of reconfigurable waveplates based on liquid crystal phase retarders in combination with a polarisation beam splitter, which can also serve as an output coupler \cite{OE_Radnatarov_2013, OE_Winters_2017, NC_Wu_2022}. In the latter case, the voltages applied to the liquid crystal elements can be precisely controlled via external drivers, enabling rapid and programmable tuning of the polarisation state, thus facilitating the efficient exploration of a broad parameter space.

The pulsating (breathing) mode-locking regime typically emerges below the pump threshold for conventional soliton mode locking in normal-dispersion cavities \cite{SA_Peng_2019}, or at higher pump powers beyond the stability range of solitons in cavities with near-zero net dispersion  \cite{PRL_Xian_2020, OL_Huang_2021, PRL_Wu_2023, PRL_Kang_2024}.

\subsubsection{Diagnostic tools}
\label{sec:Diagnostics}
\begin{figure}
\centering
{
\resizebox*{13.cm}{!}
{\includegraphics{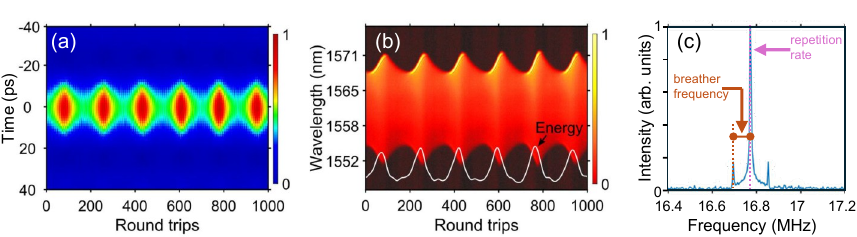}}}\hspace{5pt}
\caption{{\bf Typical properties of a breathing soliton with a long pulsation period}, observed in a laser cavity operating at normal average dispersion. The cavity repetition rate is 16.765$\,$MHz.
(a) Temporal evolution of the intensity relative to the average over successive cavity round-trips, recorded using a 50-GHz photodiode with a 20-ps response time and a 33-GHz bandwidth oscilloscope operating at an  80-GSa/s sampling rate.
(b) DFT measurement of single-shot spectra over consecutive round-trips; the white curve indicates the pulse energy evolution. The accumulated dispersion is $-1200,\mathrm{ps/nm}$, yielding a spectral resolution of 0.025$\,$nm.
(c) RF spectrum obtained by Fourier transformation of the photodiode signal. Data adapted from \cite{LPR_Wu_2022}, acquired following laser optimisation via a GA.} 
\label{fig:Experimental_breather}
\end{figure}

The characterisation of pulses emitted by a breather laser differs fundamentally from that of a conventional mode-locked laser, where identical pulses are emitted on every cavity round-trip. In such conventional systems, both the optical spectrum and the temporal profile can be readily accessed using standard tools such as optical spectrum analysers or time-domain techniques like optical autocorrelation. The averaged signals produced by these methods faithfully represent any individual pulse. This is not the case for breathing solitons. In these systems, the pulse energy fluctuates from one round-trip to the next. As a result, the electrical signal detected by a photodiode exhibits intensity modulations that reveal period-multiplication dynamics---phenomena observed in non-fibre lasers \cite{OL_Sucha_1995,OC_Wing_1999,OL_Fernandez_2004} as well as in fibre laser systems \cite{PRE_SotoCrespo_2004, OE_Zhao_2004,OC_Zhao_2007}. Figure~\ref{fig:Experimental_breather}(a) illustrates such oscillatory behaviour, showing the round-trip-resolved dynamics of a long-period pulsating soliton recorded using a high-bandwidth photodiode and oscilloscope. This measurement employs the spatio-temporal dynamics methodology \cite{AS_Sugavanam_2016}, which consists of recording a long, real-time intensity trace $I(t)$, segmenting it into time windows aligned with cavity round-trips, and assembling these segments into a false-colour contour plot. The resulting spatiotemporal map $I(t,z)$
captures the evolution of pulse dynamics over successive round-trips, where the slow coordinate $z$ indexes the round-trip number.

However, current limitations in optoelectronic detection bandwidth restrict access to the full ultrashort temporal features of breathing pulses and may, for example, obscure two closely spaced temporal components. This challenge is overcome using a method known as {\itshape time-stretch dispersive Fourier transform (DFT)}, which enables real-time visualisation of the spectral properties of individual pulses. DFT has become a breakthrough technique in the characterisation of ultrafast events and is now widely adopted by the scientific community \cite{NP_Goda_2013, NP_Mahjoubfar_2017, RPP_Wang_2020, APX_Godin}.
The principle of DFT is conceptually analogous to the far-field regime in paraxial diffraction. It relies on the fact that, under sufficient dispersion, the temporal intensity profile of a pulse becomes a stretched replica of its optical spectrum \cite{OL_Jannson_1983}. This linear spectral-to-temporal mapping---typically achieved via propagation through several kilometers of highly dispersive fibre---translates the spectral information into the time domain, enabling direct recording with a fast photodiode and a high-bandwidth oscilloscope. This property has been widely exploited, particularly in the characterisation of extreme events and transient laser dynamics \cite{NP_Ryczkowski_2018, NP_Herink_2016, PRL_Liu_2018, LPR_Peng_2018, SR_Lapre_2019, OLT_Yuan_2024}.
Figure~\ref{fig:Experimental_breather}(b) presents the round-trip-resolved spectral evolution of the same breathing soliton shown in Fig.~\ref{fig:Experimental_breather}(a), captured using the DFT technique. The data clearly show that periodic variations in pulse energy---associated with oscillations in peak amplitude and pulse width in the temporal domain---are synchronised with spectral breathing, i.e., periodic stretching and compression of the optical spectrum.

To obtain a more detailed, real-time full-field picture of the dynamics, DFT can be combined with time-lens techniques that magnify fine temporal structures, making them compatible with the bandwidth limits of available optoelectronics \cite{OE_Wei_2017,NP_Ryczkowski_2018, OL_Zhang_2020}. DFT is also highly effective for characterising doublets of closely bound ultrashort pulses, which manifest as sinusoidal modulations in the spectral envelope \cite{PRL_Krupa_2017, SR_Lapre_2019, OE_Wang_2019, SA_Peng_2019, LPL_Lu_2023, OL_Wang_2018}. From these spectral modulations, both the temporal delay and relative phase between the constituent pulses can be extracted. When the separation between pulses exceeds several nanoseconds, the optical spectrum of each pulse can even be individually resolved \cite{SA_Peng_2019, SR_Chen_2025, OLT_Li_2024, OC_Li_2025}. Moreover, DFT is well suited for analysing bi-chromatic structures \cite{PRL_Cui_2023}. Furthermore, it has enabled the direct observation of the birth-to-annihilation dynamics of dissipative Kerr cavity solitons in coherently driven Kerr resonators \cite{ChOL_Xu_2024}, and, when combined with time-lens techniques, has resolved the dynamics of dissipative Talbot solitons in synchronised multicolour fibre lasers \cite{SA_Zhang_2024}.

The spectral resolution achieved with DFT allows detection of sharp spectral features such as Kelly sidebands, signatures of four-wave mixing \cite{PJ_Wang_2017, SR_Chen_2025} or interactions with dispersive waves \cite{PTL_Chang_2025}. It can also reveal periodic shifts in the central wavelength of the breathing soliton \cite{LPR_Peng_2021, OLT_Li_2022}, and has proven essential for detecting otherwise invisible breathing behaviour that is not discernible from pulse energy measurements alone \cite{LPR_Liu_2020}. Today, DFT is such an indispensable tool in the study of oscillatory dynamics that it is difficult to imagine an experimental study in this field without it \cite{RPP_Wang_2020,JPB_Han_2022}.

In conventional mode-locked lasers, the radiofrequency (RF) spectrum primarily provides information about the laser’s repetition rate $f_{\mathrm{r}}$, as well as its noise and timing jitter. In contrast, breather lasers exhibit an additional, defining feature: a pair of sidebands appears symmetrically around the repetition frequency, located at a distance equal to the breathing frequency $f_{\mathrm{b}}$
(see, e.g., Fig.~\ref{fig:Experimental_breather}(c)) \cite{LPR_Wu_2022, OLT_Li_2022, LP_Lu_2023, LP_Fu_2024}. RF measurements can also reveal the coexistence of multiple breathing frequencies, which give rise to trampoline-like dynamics \cite{OE_Chen_2020, PRA_Du_2021}. Notably, the sharpness of these spectral features distinguishes this behaviour from Q-switching instabilities.
As such, the RF spectrum serves not only as a diagnostic tool but also as a valuable input for automated cavity control algorithms. A more detailed physical interpretation of the RF components in breather lasers will be presented in Sections~\ref{sec:Synchronisation} and~\ref{sec:Complexes}, where their connection to synchronisation mechanisms will be established. Finally, high-precision frequency counters (cymometers) may be used to resolve the RF components and evaluate their long-term stability.

\subsubsection{Control via genetic algorithms}
\label{sec:GA}
Despite their fundamental importance, breathing solitons have received comparatively little experimental attention---although this situation is rapidly evolving---primarily because their intrinsic oscillations are difficult to characterise and reproduce in a controlled, repeatable manner. As mentioned in the Introduction, fibre-laser cavities offer numerous tunable degrees of freedom \cite{Nano_Mao_2025}, including pump power and small-signal gain \cite{OE_Shen_2025, APS_Hou_2025}, output-coupler extraction ratio \cite{OLT_Han_2022}, cavity length and average dispersion \cite{AS_Boscolo_2015}, as well as temporally modulated losses implemented via acousto- \cite{CPB_Yue_2009, COL_Liu_2009} or electro-optic modulators \cite{Optica_Zou_2022, OL_Feng_2025}. These parameters define a vast and complex optimisation landscape.
In practice, the degree of freedom most commonly exploited is the effective saturable absorber. In systems based on nonlinear polarisation evolution, this corresponds to tuning the intra-cavity state of polarisation, facilitated by the active components described in Section~\ref{sec:laser_setup}.

The recent implementation of evolutionary and GAs has overcome a major bottleneck by automating exploration of this high-dimensional parameter space. GAs now enable reproducible access to complex operating regimes---such as breathing dynamics---that previously required laborious empirical tuning and whose admissible parameter window is typically narrower than that required for conventional stationary mode locking. The GA approach is particularly well-suited for the global optimisation of user-defined targets arising from complex nonlinear interactions---for instance, those involved in supercontinuum generation \cite{FP_Hoang_2022, OL_Hary_2023}. In photonic cavities, early successes include the automated identification of stationary mode-locked regimes, such as single-pulse \cite{JOSA_B_Andral_2016, SR_Woodward_2016, OFT_Han_2024}, dual-pulse \cite{OL_Girardot_2022}, and harmonic mode-locking states \cite{Optica_Pu_2019}. More recent developments have pushed these so-called {\itshape smart lasers} further, demonstrating their ability to self-tune to highly dynamic and complex regimes, including ultra-broadband noise-like emission \cite{SR_Lapre_2023} and the generation of super rogue waves \cite{LPR_Wu_2023}.

\begin{figure}
\centering
{
\resizebox*{13.cm}{!}
{\includegraphics{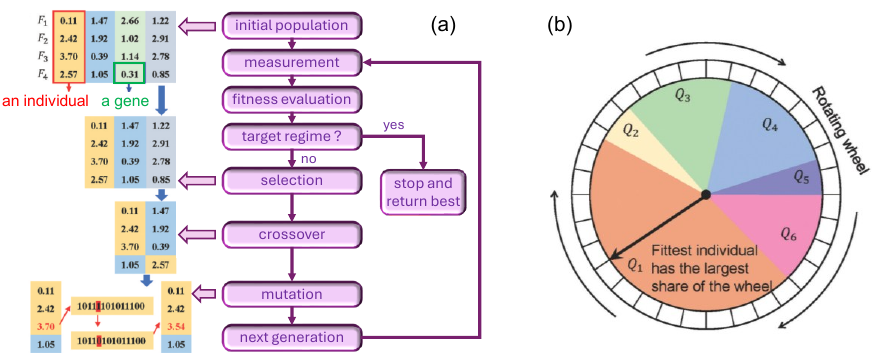}}}\hspace{5pt}
\caption{{\bf Genetic algorithm principles.} (a) Flow chart of the algorithm; (b) ``Roulette wheel'' selection diagram. Results adapted from \cite{LPR_Wu_2022}.} \label{fig:Genetic_algorithm}
\end{figure}

GAs emulate Darwinian evolution, advancing only the fittest individuals through successive generations \cite{SA_Holland_1992, Mitchell_1998} (see Fig. \ref{fig:Genetic_algorithm}(a)). In the present context, an {\itshape individual} corresponds to a particular laser operating regime, uniquely specified by the set of control voltages applied to the intracavity polarisation controller \cite{LPR_Wu_2022}; these voltages serve as that individual’s {\itshape genes}.
The optimisation begins with an initial population whose gene values are random. For every individual, the laser output is evaluated with a user-defined merit (fitness) function, which assigns a quantitative score. A new generation is then created by breeding individuals from the previous generation, with the probability of an individual being selected as a parent weighted by its score. Breeding consists of {\itshape crossover}---gene exchange between two parents---to produce two offspring, followed by {\itshape mutation}, which randomly perturbs individual genes to maintain genetic diversity.
Selection strategy is regime-dependent. For stationary mode-locking, an {\itshape elitist} scheme---in which the highest-scoring individuals are cloned directly into the next generation---rapidly preserves desirable traits. By contrast, for breather mode locking, a {\itshape roulette-wheel} scheme, which assigns selection probabilities proportional to fitness yet still explores lower-ranking individuals, proves more effective \cite{CJL_Wu_2023} (see Fig. \ref{fig:Genetic_algorithm}(b)). Generations iterate until the population converges, yielding the individual (laser setting) that maximises the fitness function, i.e., the desired operating state.

The merit function is the linchpin of any self-tuning scheme: it must increase monotonically as the laser approaches the target state. Simple metrics---such as the peak height at the cavity repetition frequency or the detected pulse count---are effective for locating stationary mode-locked regimes and can be adapted to identify basic breather operation \cite{OE_Zheng_2021}. To reach breather states with prescribed attributes, however, it is advantageous to exploit finer details of the RF spectrum \cite{LPR_Wu_2022}. For an automatically optimised, self-starting breather regime, we define the composite merit function
\begin{equation}
F_{\mathrm{merit}} = \alpha F_{\mathrm{ml}} + \beta F_{\mathrm{b}},\quad F_{\mathrm{b}}=1-\sum_{f_{\mathrm{r}}-\Delta}^{f_{\mathrm{r}}+\Delta}I(f)/\sum_{f_{-1}}^{f_{+1}}I(f),
\label{eq:merit}
\end{equation}
where $F_{\mathrm{ml}}$ quantifies the quality of mode locking and is taken as the average pulse intensity \cite{SR_Woodward_2016, OE_Zheng_2021}. Its role is to penalise operating states---such as noise-like pulsing or relaxation oscillations---that can mimic breather spectra. The second term $F_{\mathrm{b}}$ discriminates between stationary and breathing operation. In a breather, the modulation frequency $f_{\mathrm{b}}$ appears as symmetric sidebands  $f_{\pm 1}$ around the cavity repetition frequency $f_{\mathrm{r}}$ [$f_{\mathrm{b}}=\abs{f_{\pm 1}-f_{\mathrm{r}}}$; see Fig.\ \ref{fig:Experimental_breather}(c)]; in a stationary mode-locked state, these sidebands are absent. Accordingly, $F_{\mathrm{b}}$ is constructed from the ratio of the spectral power at $f_{\mathrm{r}}$ (summed over a window of width $2\Delta$)  to the power contained in the sideband region spanning $f_{-1}$ to $f_{+1}$. The weighting coefficients $\alpha$ and $\beta$ are chosen empirically. By augmenting $F_{\mathrm{merit}}$ with additional terms, we have obtained fine control over breather attributes such as the breathing ratio and period \cite{LPR_Wu_2022}. When the pump power is set high enough to favour multi-pulse self-starting, the same composite merit function---followed by a pulse-count constraint---enables the GA to stabilise breather molecular complexes with a user-specified number of constituent breathers, as demonstrated in Section \ref{sec:GA}.
A further refinement exploits the defining hallmark of {\itshape frequency-locked} breathers, namely the high signal-to-noise ratio (SNR) at the breathing sidebands [Fig.\ \ref{FrequencyLocking}(a3-a4)]. Incorporating this SNR metric into $F_{\mathrm{merit}}$ proved essential for directing the GA to laser states that exhibit precise frequency locking, thereby allowing systematic exploration and tailoring of the Farey-tree hierarchy of locked ratios reported in Ref.\ \cite{NC_Wu_2022} and discussed in Section \ref{FareyTree}.

This area is undergoing rapid development on both the algorithmic and hardware fronts \cite{ML_Sun_2020, IPT_Liu_2024, AOP_DiLauro_2025}. Emerging control strategies---most notably those that integrate neural-network (NN) \cite{SR_Han_2024, IPT_Zhang_2024} surrogates with evolutionary search---are markedly improving optimisation speed and accuracy, even as the dimensionality of the parameter space grows. These advances are now enabling real-time tuning of increasingly intricate mode-locking regimes that were previously beyond practical reach.

\subsection{Laser model}
In this section, we present and discuss the two principal physical models that have underpinned the conceptualisation and characterisation of pulsating solitons as a distinct class of nonlinear dissipative structures. While these models remain foundational for physical insight, the rapid development of machine learning (ML) has recently introduced alternative approaches, particularly those based on NN architectures of varying complexity \cite{LPR_Pu_2023, LPR_Si_2024, OC_Zhang_2025}. Although these data-driven models offer promising avenues---especially due to their high-speed simulation capabilities after training---they typically require a substantial dataset for training, often generated numerically. Moreover, their ``black-box'' nature can obscure the underlying physics, making it difficult to interpret the contribution of individual components. Despite these limitations, such approaches are increasingly valuable for complementing traditional modelling strategies, particularly when navigating high-dimensional or experimentally inaccessible regimes.

\subsubsection{Master-equation approach based on the cubic-quintic Ginzburg-Landau equation}

To investigate the dynamics of fibre lasers and assess the generality of experimentally observed behaviours, a widely adopted approach is to perform numerical simulations based on the master equation formalism---one of the foundational methods in the theory of passively mode-locked lasers. Originally introduced by Haus \cite{JAP_Haus_1975}, this framework has evolved into the cubic–quintic Ginzburg–Landau equation (CQGLE), which is often considered the minimal-complexity model that still supports soliton solutions. In its standard form, the CQGLE is expressed as
\begin{equation}
    i\Psi_{\xi} + \frac{D}{2} \Psi_{\tau\tau} + \abs{\Psi}^2 \Psi + \eta \abs{\Psi}^4 \Psi =
    i\Theta \Psi + i\epsilon \abs{\Psi}^2 \Psi + i\beta \Psi_{\tau\tau} + i\mu \abs{\Psi}^4 \Psi,
\label{CQGLE}
\end{equation}
where $\Psi$ is the normalised complex envelope of the optical field, $\tau$ denotes the normalised time in a reference frame moving at the group velocity, and $\xi$ represents the propagation distance along the unfolded cavity. The dimensionless temporal and spatial coordinates are scaled by a characteristic pulse duration $T_0$ and the dispersion length $L_{\mathrm{D}}=T_0^2/\abs{\overline{\beta}_2}$, respectively, where $\overline{\beta}_2$ is the path-averaged group-velocity dispersion of the cavity. On the left-hand side of Eq.\ \ref{CQGLE}, the parameter $D=-\mathrm{sgn}(\overline{\beta}_2)$ indicates the sign of the dispersion regime, and $\eta$ quantifies the quintic nonlinear refractive index contribution. The right-hand side contains the dissipative terms: $\Theta$
represents the net linear gain or loss, $\beta$ is the gain bandwidth parameter, and $\epsilon$ and $\mu$ are the cubic and quintic gain/loss coefficients, respectively.

The master-equation framework is invaluable for discerning the spectrum of nonlinear structures that can emerge in an optical cavity \cite{NP_Grelu_2012}. Beyond stationary dissipative solitons, the CQGLE predicts {\itshape pulsating} solitons---periodic attractors that occupy an intermediate state between stationary behaviour and chaos \cite{PRE_Akhmediev_2001, PRE_SotoCrespo_2004}. In phase space, stationary solitons correspond to fixed points, whereas pulsating solitons trace limit cycles \cite{PLA_Tsoy_2005}. The CQGLE also reproduces extreme breathing dynamics \cite{PRE_Chang_2015}, eruptive structures \cite{PRL_Sotocrespo_2000}, soliton pairs (molecules) exhibiting vibrational oscillations, and compound breathing states \cite{SA_Peng_2019, LPR_Peng_2021, OL_Du_2022}. Moreover, it captures routes to chaos via successive bifurcations \cite{PRE_Akhmediev_2001, PRE_SotoCrespo_2004}. Because key physical parameters such as dispersion, gain, and saturation enter explicitly, the model readily accommodates higher-order linear and nonlinear effects \cite{OC_Song_2005, PRE_Uzunov_2018, OE_He_2018, CSF_Nana_2023}. Although the CQGLE is most often integrated numerically, reduced-order solutions can be obtained via moment methods \cite{PLA_Tsoy_2005, CSF_Nana_2023, JLT_Li_2025} or variational techniques employing trial functions \cite{IET_Ferreira_2018}, reinforcing its versatility as a universal model for dissipative systems.

\subsubsection{Lumped model based on the generalised nonlinear Schr\"odinger equation}
\label{LumpedModel}
Despite the considerable success of the CQGLE in providing a qualitative and descriptive understanding of the main families of solutions that can arise in fibre lasers, it remains challenging to apply this approach when accurate quantitative agreement with experimental observations is required. Consequently, a more reliable modelling framework for optimising fibre cavity design has progressively gained attention. This challenge has been addressed through a {\itshape lumped modelling} approach, wherein each section of the fibre cavity is modelled individually. Unlike the CQGLE framework---whose parameters are difficult to relate directly to the physical characteristics of specific cavity segments---the lumped model enables straightforward incorporation of experimentally measured values, thereby improving the fidelity of simulations. Moreover, it offers valuable physical insight into the nonlinear pulse dynamics occurring within each segment or component of the cavity.

As previously noted, by focusing in this study on pulse regimes that do not involve periodic energy exchange between polarisation modes, a scalar approximation is sufficient to describe light propagation in the fibre. The generalised NLSE is used to model the pulse evolution in each fibre segment \cite{JSTQE_Haus_2000}: 
\begin{equation}
    \psi_z = -\frac{i\beta_2}{2}\psi_{tt} + i\gamma \abs{\psi}^2 \psi + \frac{g}{2} \left( \psi + \frac{1}{\Omega^2} \psi_{tt} \right),
    \label{eq:GNLSE}
\end{equation}
where $\psi = \psi(z,t)$ denotes the slowly varying electric field envelope, $t$ is the retarded time, and $z$ is the longitudinal propagation coordinate. The coefficients $\beta_2$ and $\gamma$ represent second-order dispersion and Kerr nonlinearity, respectively. The dissipative terms account for linear gain as well as a parabolic approximation of the gain spectral profile, with bandwidth $\Omega$.
The gain is modelled as a saturable function given by $g(z) = g_0 \exp\left(-\frac{E_{\mathrm{p}}}{E_{\mathrm{sat}}}\right)$, where $g_0$ is the small-signal gain (nonzero only in the gain fibre), $E_{\mathrm{p}}(z) = \int \mathrm{d}t\, \abs{\psi}^2$ is the pulse energy, and $E_{\mathrm{sat}}$ is the saturation energy determined by the pump power.
More advanced gain dynamics can also be implemented, particularly to account for the population dynamics of the active ions in the doped fibre \cite{PRL_Cui_2023, OLT_Yuan_2024}, and to provide insight into Q-switching-related regimes.  

The NPR-based mode-locking mechanism can be modelled using an instantaneous and monotonically increasing nonlinear transfer function applied to the field amplitude:
\begin{equation}
    T = \sqrt{1 - q_0 - q_{\mathrm{m}} \left/ \left[ 1 + \frac{P(t)}{P_{\mathrm{sat}}} \right] \right.},
    \label{eq:NPE}
\end{equation}
where $q_0$ denotes the unsaturated loss of the absorber, $q_{\mathrm{m}}$ is the saturable loss (modulation depth), $P(z,t) = \abs{\psi(z,t)}^2$ is the instantaneous pulse power, and $P_{\mathrm{sat}}$ is the saturation power.
Linear losses---accounting for intrinsic fibre losses and output coupling---are typically imposed after the passive fibre segments. 

The numerical model is generally solved using a symmetric split-step Fourier method \cite{book_Agrawal}. To improve convergence towards the nonlinear steady-state structure, simulations often begin with a Gaussian-like initial condition rather than random noise.
It is worth noting that the inclusion of higher-order dispersive terms, such as fourth-order dispersion, has revealed the potential existence of pulsating quartic solitons \cite{CSF_Yang_2023, CSF_Luo_2025, OE_Wu_2025, LPR_Wang_2025B}.

\section{Synchronisation dynamics}
\label{sec:Synchronisation}

First identified by Christiaan Huygens in 1665 \cite{huygens_1665}, frequency locking---or synchronisation---is the process by which coupled nonlinear oscillators adjust their frequencies to match or maintain a rational ratio (known as the winding number). This phenomenon is ubiquitous across both natural and engineered systems, appearing in contexts as diverse as biological clocks, chemical reactions, mechanical and electrical oscillators, and lasers, to name just a few well-known examples  \cite{Pikovsky_2001}. It underpins a broad range of technologies, including telecommunications, global navigation systems, and biomedical instrumentation.  Recently, laser cavities and microresonators operating in the breathing-soliton regime have emerged as a compelling platform for investigating synchronisation dynamics within a single physical system \cite{PRL_Cole_2019, PRL_Xian_2020, NC_Wu_2022, PRL_Wu_2023}. In these systems, harmonics of the breathing frequency $f_{\mathrm{b}}$ can lock to the cavity repetition frequency $f_{\mathrm{r}}$, driven by nonlinear coupling and competition between the intrinsic system frequencies, with $f_{\mathrm{r}}$ acting as the master and $f_{\mathrm{b}}$ as the slave. This self-synchronisation mechanism obviates the need for external modulation or auxiliary resonators, offering a compact and inherently robust approach to frequency locking. In the following, we demonstrate that breathing-soliton lasers serve as enabling platform for  the emergence of intricate synchronisation phenomena and the manifestation of self-similar fractal dynamics in nonlinear systems.

\subsection{Different breather states}
\begin{figure}
\centering
{\resizebox*{13.5cm}{!}{\includegraphics{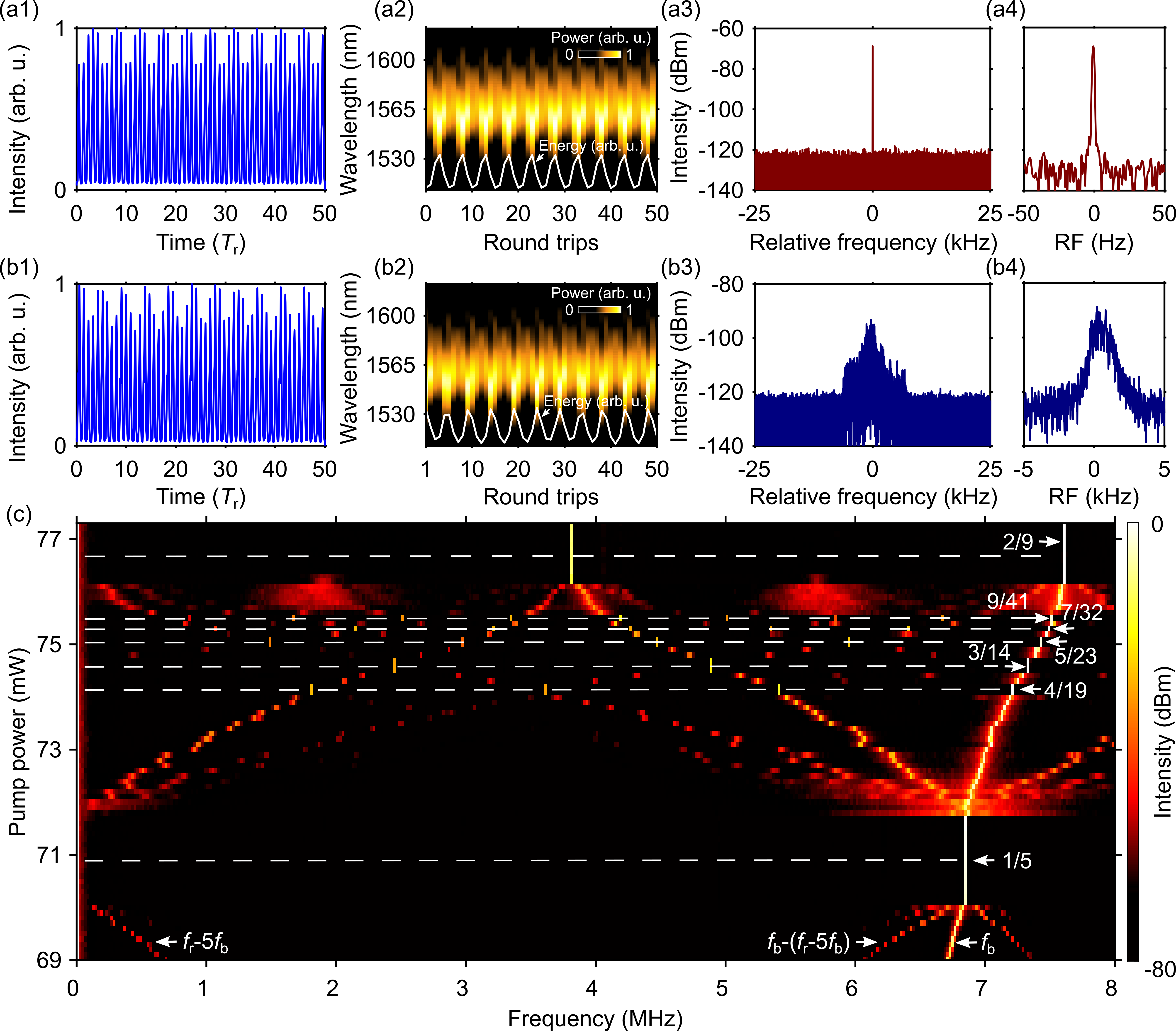}}}  
\caption{\textbf{(a,\,b) Experimental characterisation of synchronised and unsynchronised breathing-soliton states.} 
(a1,\,b1) Photodetected dispersive Fourier transform (DFT) signals captured over consecutive cavity roundtrips ($T_{\mathrm{r}}$ denotes the roundtrip time).
(a2,\,b2) Corresponding single-shot DFT spectra; white curves trace the energy evolution.
(a3-a4,\,b3–b4) Associated RF spectral measurements. The synchronised state (a3–a4) shows a single-mode oscillation at the subharmonic breathing frequency over spans of 50$\,$kHz and 100$\,$Hz. In contrast, the unsynchronised state (b3–b4) exhibits unstable multimode oscillation of a non-subharmonic breathing frequency over 50-kHz and 10-kHz spans. The reference frequency corresponds to one-fifth of the fundamental repetition rate.
{\bf (c) Experimental observation of frequency locking:} RF spectrum of the laser output versus pump power, showing the sequential emergence of rational winding numbers. Adapted from \cite{NC_Wu_2022}.
} \label{FrequencyLocking}
\end{figure}

As discussed in the Introduction, mode-locked lasers can support not only stationary dissipative solitons but also breathing solitons, depending primarily on the pump strength and the intra-cavity polarisation state, the latter effectively modulating the cavity loss \cite{PRL_Xian_2020,NC_Wu_2022,PRL_Wu_2023,SA_Peng_2019,LPR_Peng_2021,LPR_Wu_2022,LPR_Wang_2023,PRL_Cui_2023}. The transition from stationary to breathing solitons is associated with a Hopf bifurcation---a widely observed dynamical instability marking the onset of periodic behaviour in nonlinear systems. Figures \ref{FrequencyLocking}(a) and \ref{FrequencyLocking}(b) illustrate two distinct breather regimes that emerge at closely spaced pump powers: a subharmonically synchronised (frequency-locked) breather state, and an unsynchronised (quasi-periodic) state \cite{NC_Wu_2022}. In the synchronised regime [Figs.\ \ref{FrequencyLocking}(a1-2)], the photo-detected signal after time stretching, optical spectrum, and pulse energy exhibit strictly periodic variations with a well-defined period---5 cavity roundtrips in this case---indicative of subharmonic entrainment.
By contrast, the quasi-periodic regime [Figs.\ \ref{FrequencyLocking}(b1-2)] displays degraded temporal and spectral periodicities. The most prominent distinction between the two regimes is evident in the corresponding RF spectra [Figs.\ \ref{FrequencyLocking}(a3-4,\,b3-4)]. The synchronised state exhibits a sharp breathing frequency component with narrow linewidth (0.5$\,$Hz) and a high SNR, precisely located at one-fifth of the cavity repetition frequency, yielding a rational winding number $f_{\mathrm{b}}/f_{\mathrm{r}}=1/5$. Conversely, the unsynchronised state is characterised by a broadened, noisy breathing frequency that deviates from this rational subharmonic. Direct time-resolved measurements of the breathing frequency using a cymometer reveal marked disparities in frequency stability between the two regimes \cite{NC_Wu_2022}. As explored further in the following subsection, the locked breathing frequency remains robust over a range of pump power values, indicating the presence of a synchronisation plateau.

The distinct characteristics of the two breather states are also evident in the pump-power-resolved RF spectrum shown in Fig.\ \ref{FrequencyLocking}(c), which captures the onset and evolution of frequency locking. At lower pump powers, three dominant RF components are observed: the breathing frequency, the difference frequency between $f_{\mathrm{r}}$ and the 5th harmonic of $f_{\mathrm{b}}$ ($f_{\mathrm{r}}-5f_{\mathrm{b}}$), and the difference frequency between the first two ($6f_{\mathrm{b}}-f_{\mathrm{r}}$).
 As $f_{\mathrm{r}}-5f_{\mathrm{b}}$ approaches zero, the system undergoes a transition to a locked state with a winding number of $1/5$. 
Further increases in pump power induce redshifts in this winding number, giving rise to a sequence of rational ratios and indicating the emergence of a devil's staircase structure characteristic of frequency-locking transitions in nonlinear systems, as discussed in the next subsection. Notably, the observation of breather frequency locking requires the net cavity dispersion to be close to zero; such dynamics are absent when the laser operates under moderate or large normal dispersion \cite{SA_Peng_2019}. In the latter case, the breathing frequency is comparatively slow, leading to RF spectra densely packed around $f_{\mathrm{r}}$ (see, e.g., Fig.\ \ref{fig:Experimental_breather}(c)) and to sidebands that are not located exactly at subharmonics of $f_{\mathrm{r}}$. This reflects that the breathing oscillation is not readily commensurate with the cavity round-trip time. By contrast, near-zero dispersion yields much faster breathing oscillations---by an order of magnitude or more (cf.\ Section \ref{sec:Diagnostics})---which are easily subharmonically related to $f_{\mathrm{r}}$ and therefore capable of frequency locking. This distinction highlights the fundamentally different physical mechanisms underlying the two breathing-soliton regimes \cite{Submitted_Wu_2025}.

Our further investigations \cite{PRL_Wu_2023} have revealed the existence of an intermediate state between the synchronised and unsynchronised phases of breather structures, which we refer to as the {\itshape modulated subharmonic state}. To the best of our knowledge, this regime has not been previously observed in nonlinear systems. A characterisation of this state---distinguished by a self-modulation of the subharmonically synchronised breather oscillations---is presented in Section \ref{sec:SynchDesynch}.

\subsection{Farey tree and devil's staircase}
\label{FareyTree}

Frequency locking has been studied across a wide range of physical systems, including  charge-density waves \cite{RMP_Gruner_1988}, Josephson junctions \cite{PRL_Shapiro_1963}, and the Van der Pol oscillator \cite{Nature_VanDerPol_1927}, among others \cite{Nature_Zhang_2017}. The distribution of frequency-locked states in parameter space---exhibiting the fractal structure known as the devil’s staircase \cite{PRL_Jensen_1983, PT_Bak_1986}---can be understood through the number-theoretic framework of Farey trees \cite{PRB_Bak_1980, PRL_Brown_1984, Science_Jin_1994, NC_Kuroda_2020}. The Farey tree is a hierarchical sequence of rational numbers constructed via the Farey-sum (or mediant) operation, denoted by $\oplus$: given two adjacent fractions, $\frac{m}{n}$ and $\frac{p}{q}$, a new fraction is generated at the next level of the tree by summing the numerators and denominators separately, yielding $\frac{m}{n} \oplus \frac{p}{q} = \frac{m + p}{n + q}$. In the context of nonlinear dynamics, the Farey tree provides a framework for understanding the local organisation of two-frequency resonances. The physical motivation for its use lies in the observation that, between two adjacent resonances, the Farey fraction---with the smallest denominator---is typically the dominant resonance in that interval. This hierarchical structure gives rise to a devil’s staircase curve composed of an infinite number of plateaux, exhibiting characteristic self-similarity and fractal geometry. 

In optics, frequency-locking phenomena have been extensively studied in externally modulated semiconductor lasers \cite{APL_Winful_1986, PRL_Baums_1989, PRL_Wunsche_2005, OE_Shortiss_2019, NC_Greilich_2025}, where tuning the modulation frequency allows direct observation of the Farey hierarchy and the devil’s staircase structure \cite{PRL_Baums_1989}. Frequency locking has also been observed in a range of other systems, including coupled or externally driven Kerr resonators \cite{NP_Jang_2018, Optica_Xu_2020}, coupled semiconductor laser oscillators \cite{PRL_Thevenin_2011}, fibre lasers with externally modulated loss \cite{CPB_Yue_2009, COL_Liu_2009} \textcolor{red}{or gain \cite{JLT_Li_2025}}, solid-state lasers operating in dual-mode regimes \cite{PRA_Otsuka_2000}, and in the generation of soliton-pair molecules in solid-state lasers under external modulation \cite{NP_Kurtz_2020}. In all of these cases, synchronisation is driven by an auxiliary oscillator or an external modulation source that introduces a distinct characteristic frequency into the system.
In contrast, optical resonators supporting breathing solitons  inherently exhibit two characteristic frequencies, thereby offering a fundamentally different platform for studying frequency locking without external forcing. 

\begin{figure}
\centering
{\resizebox*{14.2cm}{!}{\includegraphics{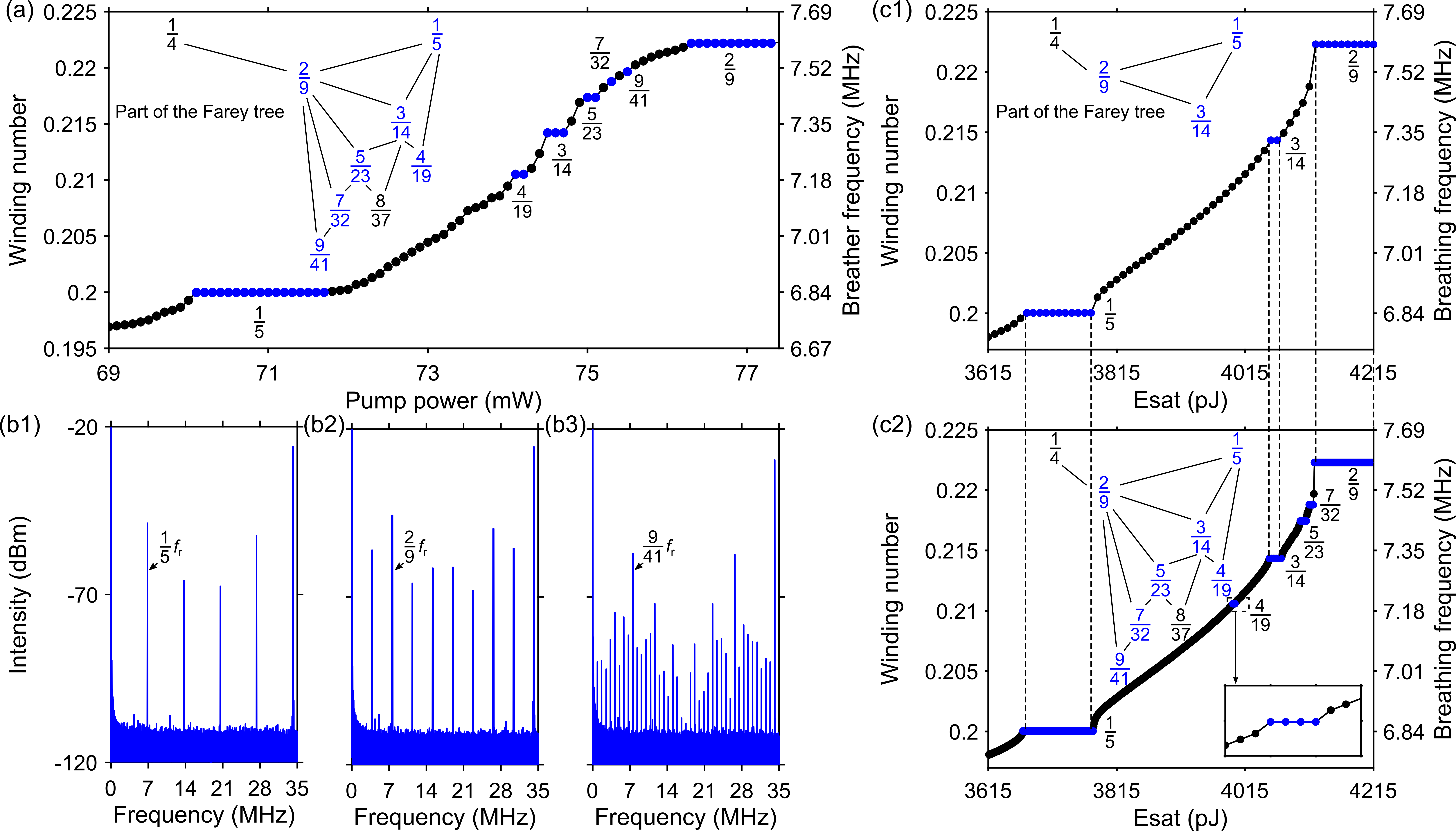}}}
\caption{{\bf Farey tree and devil's staircase}. (a) Measured breathing frequency (winding number) plotted as a function of pump power. The inset shows the relevant portion of the Farey tree, with the observed winding numbers highlighted in blue. (b) RF spectra corresponding to frequency-locked states with winding numbers 1/5, 2/9, and 9/41, respectively. In each case, a set of equidistant spectral lines emerges within the frequency span defined by the cavity repetition rate $f_{\mathrm{r}}=34.2\,$MHz. 
(c) Simulated breathing frequency (winding number) as a function of the gain saturation energy, varied with step sizes of 10$\,$pJ and 1$\,$pJ, respectively. The finer step reveals additional plateaux, indicating a fractal structure in the frequency-locking behaviour. Insets display the relevant sections of the Farey tree with the observed Farey fractions. Adapted from \cite{NC_Wu_2022}.} \label{Fareytree}
\end{figure}

In \cite{NC_Wu_2022}, by systematically exploring transitions between different breather states in a laser cavity accessible via pump-power tuning, we reported, for the first time, a high-order Farey tree hierarchy of frequency-locked states. Figure \ref{Fareytree}(a) shows a representative measurement of the breathing frequency as a function of pump power, starting from the region corresponding to a winding number of 1/5 (see Fig.\ \ref{FrequencyLocking}). The data reveals a characteristic devil’s staircase structure, with distinct plateaux. The frequencies corresponding to the plateaux can be associated with rational winding numbers, as identified through the analysis of the RF spectra [Fig.\ \ref{Fareytree}(b)]. In each frequency-locked state, the RF spectrum displays a finite number $n$ of equally spaced sidebands below the cavity repetition rate $f_{\mathrm{r}}$. The most prominent line corresponds to the breathing frequency $f_{\mathrm{b}}$, and if this line is the $m$‑th component from the low-frequency side, then the associated winding number is given by $f_{\mathrm{b}}/f_{\mathrm{r}}=m/n$. Remarkably, the winding numbers emerge sequentially across the pump power axis in the order prescribed by the Farey tree [inset of Fig.\ \ref{Fareytree}(a)], with the width of each plateau  inversely correlated with the level at which the corresponding fraction $m/n$ appears in the Farey hierarchy. The gaps in pump power between adjacent steps (plateaux) correspond to regions of quasi-periodic breather oscillations, similar to the example shown in Fig.\ \ref{FrequencyLocking}(b). The fractal dimension $D$ of the staircase, calculated from the distribution of these gaps \cite{PD_Hentschel_1983}, is found to be $D=0.906\pm 0.025$---closely approaching the theoretical value of 0.87 predicted for a complete devil’s staircase, as described by the circle map model \cite{PRL_Jensen_1983}, thereby highlighting the universal nature of frequency-locking dynamics in nonlinear systems governed by two competing frequencies.

A remarkable feature emerging from Fig.\ \ref{Fareytree}(b) is that the frequency-locked breather regime gives rise to the excitation of dense RF combs---e.g., 41 times denser than those produced in the standard single-pulse stationary regime in the case of Fig.\ \ref{Fareytree}(b3). Notably, the line spacing in these combs is not constrained by the cavity length and can extend into the sub-MHz range. As a stable alternative to long fibre cavities, such lasers hold strong potential for applications like high-resolution spectroscopy.

Figure \ref{Fareytree}(c) presents the corresponding numerical simulation results, showing the breathing frequency as a function of the gain saturation energy, which plays a role analogous to the pump power in the experiment (cf.\ Section \ref{LumpedModel}). The parameter is varied starting from the range corresponding to the 1/5 locked state, using two different step sizes. With finer steps, a larger number of frequency-locked plateaux emerge, confirming the fractal nature of the winding number distribution. As shown in Fig.\ \ref{Fareytree}(c2), the model successfully reproduces the same portion of the Farey tree observed in the experiment, spanning breathing frequencies from 1/5 to 2/9. Moreover, the gaps between the steps in gain saturation energy resemble those found experimentally in pump power. The fractal dimension of the gap set calculated from the model is $D = 0.873 \pm 0.009$, which is even closer to the theoretical value expected for a complete devil’s staircase---thanks to the ability of the model to use arbitrarily small increments in gain saturation energy. 

A slight change in the initial polarisation state of the laser in the experiment---or equivalently, a slight variation in the intracavity loss in the model---can trigger the emergence of Farey fractions from different branches of the Farey tree \cite{NC_Wu_2022}. Our simulations further demonstrate that the frequency-locking phenomenon is robust across a wide range of laser parameters, including net cavity dispersion and the modulation depth of the saturable absorber \cite{OC_Zhang_2023}. More recently, experiments using a figure-nine laser configuration \cite{OLT_Yuan_2025} and dispersion tuning to induce Farey-tree locking \cite{JOSAB_Ni_2024} have reinforced the robustness and universality of the fractal dynamics observed in breather lasers. These findings have also stimulated analogous investigations into self-synchronisation phenomena in Kerr resonators \cite{Chaos_Dong_2025} and fractal dynamics in terahertz quantum cascade lasers \cite{LSA_Liu_2025}.

\subsection{Complexity of Arnold tongues}
In the classical master–slave synchronisation scheme described by Adler's equation, the frequency of the slave oscillator locks to that of the master when their frequency detuning lies within a specific range \cite{Pikovsky_2001,IRE_Adler_1946}. This locking range expands with increasing coupling strength, forming a characteristic tongue-shaped region---commonly referred to as an {\itshape Arnold tongue} after mathematician Vladimir I. Arnold \cite{Arnold_1978}---in the parameter space defined by frequency detuning and coupling strength. Within this region, both frequency and phase locking are achieved. Arnold tongues serve as a fundamental tool for controlling synchronisation dynamics, which is essential for practical applications. These structures have been extensively investigated across diverse physical systems, including coupled nanomechanical oscillators \cite{Science_Shim_2007}, Kerr resonators \cite{SA_Kim_2021,PRA_Skryabin_2021}, biological oscillators \cite{NPhy_Droin_2019,SA_FontMunoz_2021}, oscillators subjected to external frequencies \cite{NC_Rodriguez_2021}, and many others \cite{PRL_Laskar_2020,PRL_Zheng_2023,PRL_Reichhardt_1999}. Additionally, Arnold tongues have been observed in the synchronisation of the internal dynamics of soliton molecules in fibre lasers under external modulation \cite{Optica_Zou_2022}. 

\begin{figure}
\centering
{\resizebox*{12cm}{!}{\includegraphics{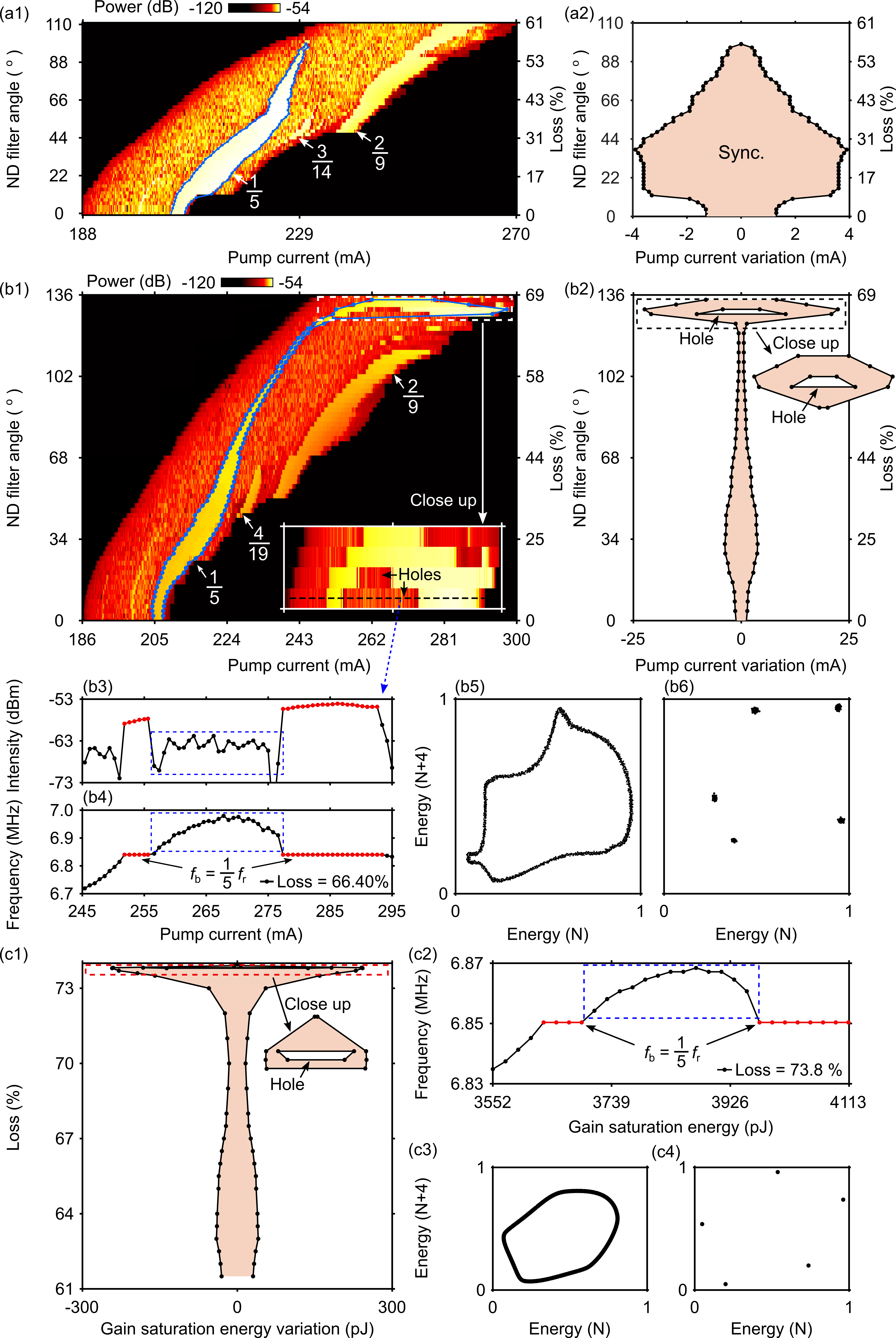}}}
\caption{{\bf Leaf-like and ray-like synchronisation regions observed in the experiment}. (a1,\,b1) Maps of the breathing frequency intensity in the parameter space defined by pump current and intra-cavity loss (the latter controlled via rotation of a neutral density filter). Regions enclosed by blue dashed contours correspond to high-intensity signals and denote the main synchronisation regions, associated with a winding number of $1/5$.
(a2,\,b2) Synchronisation regions extracted from (a1) and (b1), respectively, highlighting their distinct leaf-like and ray-like structures. For clarity, the pump current axis at each loss value is offset relative to the midpoint of the corresponding plateau in (a1,\,b1). Synchronised and unsynchronised states are indicated in pink and white, respectively.
The dashed rectangular areas in (b1) and (b2) are magnified in the corresponding insets.
(b3) Cross-section of the inset in (b1), showing a drop in breathing frequency intensity between two plateaux (red markers).
(b4) Corresponding variation of the breathing frequency with pump current.
(b5,\,b6) Poincar\'e sections for unsynchronised and synchronised states, respectively, showing the phase portraits of pulse energy at cavity roundtrip $N+4$ versus $N$. Adapted from \cite{SA_Wu_2025}.} \label{ArnoldTongues}
\end{figure}

Although Arnold tongues are considered universal features of synchronisation, pioneering theoretical studies have shown that, under sufficiently strong forcing, the locking region may cease to broaden with increasing drive strength, leading to a significant deviation from the classical tongue shape \cite{Pikovsky_2001,PR_Feudel_1997}. Instead, the region first widens and then narrows, evolving into a distinct leaf-like structure \cite{PR_Feudel_1997}. Additionally, strong forcing can give rise to {\itshape holes} within Arnold tongues---regions of quasi-periodic (unsynchronised) dynamics embedded within the synchronised domain---as demonstrated in theoretical studies of flow systems \cite{PRL_Papaioannou_2006}. Although not explicitly addressed, similar dynamical behaviour was also implied in theoretical studies of breathing solitons in optical microresonators \cite{PRL_Cole_2019}. However, experimentally accessing these nonstandard regimes remains challenging, as many real-world synchronised systems become fragile under strong forcing. In optical systems, for instance, a strong external drive from the master oscillator can disrupt the delicate coherent pulsing states of the slave oscillator \cite{NP_Jang_2018,Optica_Zou_2022,PRL_Thevenin_2011}. This phenomenon, known as {\itshape amplitude death}, refers to the complete suppression of oscillations under excessive forcing \cite{Turing_1952,Pikovsky_2001,PR_Saxena_2012}. In contrast, resonators supporting breathing solitons inherently exhibit a significant imbalance in strength, with the master oscillations ($f_{\mathrm{r}}$) being much stronger than the slave oscillations ($f_{\mathrm{b}}$)---thus creating a favorable platform for exploring abnormal synchronisation regimes. In \cite{SA_Wu_2025}, by implementing high-resolution control of intra-cavity loss (via a neutral density filter) to modulate the coupling strength between $f_{\mathrm{b}}$ and $f_{\mathrm{r}}$ in a breather laser---thereby introducing a second degree of freedom alongside the pump strength used to characterise the frequency-locking range---we demonstrated that this intrinsic asymmetry provides experimental access to synchronisation dynamics that deviate from the canonical Arnold tongue structure.
Specifically, we revealed both a leaf-like and a ray-like pattern---of which the former had previously only been studied numerically in the circle map model \cite{PR_Feudel_1997}---and experimentally observed both for the first time. In addition, we identified holes within Arnold tongues, marking the first experimental confirmation of this theoretically predicted feature \cite{PRL_Papaioannou_2006}. 

Figure \ref{ArnoldTongues}(a) presents an example of a  synchronisation pattern observed in our laser. The synchronisation region is clearly resolved in the map of breathing frequency intensity across the parameter space defined by pump current and intra-cavity loss [Fig.\ \ref{ArnoldTongues}(a1)], where it exhibits a distinct leaf-like shape, highlighted by the blue dashed outline. To better visualise this structure, the corresponding locking range is plotted as a function of pump current in Fig.\ \ref{ArnoldTongues}(a2). In addition to the main synchronisation region corresponding to a winding number of $f_{\mathrm{b}}/f_{\mathrm{r}} = 1/5$, two narrower synchronisation regions also emerge in Fig.\ \ref{ArnoldTongues}(a1), associated with winding numbers $3/14$ and $2/9$. These three ratios follow the Farey tree ordering (cf.\ Section \ref{FareyTree}). The black areas in Fig.\ \ref{ArnoldTongues}(a1) indicate either stationary soliton states or continuous-wave laser operation. 

Interestingly, the synchronisation region in Figs.\ \ref{ArnoldTongues}(b1--2)---obtained under different polarisation controller settings---exhibits a fish-ray-like structure, with a 'head' that contains a distinct 'hole', as highlighted in the insets. This hole reflects a transition from synchronised breather oscillations to an unsynchronised state, followed by re-entry into synchrony as the pump current is tuned. To illustrate this, a cross-section of the inset in Fig.\ \ref{ArnoldTongues}(b1) is shown in Fig.\ \ref{ArnoldTongues}(b3), where the drop in breathing frequency intensity corresponds to points within the hole. The breathing frequency’s variation with pump current, plotted in Fig.\ \ref{ArnoldTongues}(b4), reveals a nearly parabolic dependence between two synchronisation plateaux---a feature also predicted theoretically in breathing soliton dynamics within microresonators \cite{PRL_Cole_2019}. Such holes have been linked to quasi-periodic states \cite{PRL_Papaioannou_2006}, which is confirmed experimentally by the phase diagram in Fig.\ \ref{ArnoldTongues}(b5), characteristic of quasi-periodic dynamics \cite{Pikovsky_2001,Ott_2002,Berge_1987}. For comparison, the phase diagram of the synchronised state in Fig.\ \ref{ArnoldTongues}(b6) displays five fixed points, consistent with the winding number 1/5.

These unconventional synchronisation patterns were also reproduced in our lumped laser model within the parameter space of gain saturation energy and intra-cavity loss, with all key features---including the holes---closely mirroring those seen in the experiment \cite{SA_Wu_2025}. Notably, the high-resolution parameter sweeps required for this analysis were made computationally feasible through the use of parallel computing. 
Numerical simulations have confirmed that the transmission function of the NPR plays a central role in shaping the synchronisation regions within the system. Furthermore, these simulations have revealed the complex dependence of the locking range on intra-cavity loss. Notably, in regimes of high loss---where the slave becomes substantially weaker than the master---the foundational assumptions of Adler’s weak-injection theory \cite{IRE_Adler_1946} no longer hold.

While the physical origin of the ``holes'' appearing within otherwise continuous synchronisation regions remains unresolved, both our experimental and numerical results indicate that such features can only be observed under conditions of precise control over both linear and nonlinear intra-cavity losses. This finding may inform future experimental strategies in other nonlinear photonic systems, such as microresonators \cite{ArXiv_Moille_2025}.

\section{Breather complexes: Control and synchronisation of composite structures}
\label{sec:Complexes}

The interaction between optical solitons can give rise to compact and stable self-assembled bound states, commonly referred to as {\itshape soliton molecules} \cite{PRA_Malomed_1991,PRL_Stratmann_2005,Science_Herink_2017,PRL_Krupa_2017,PRL_Liu_2018,LPR_Peng_2018,NC_Wang_2019,Optica_Liu_2022,APN_Zou_2025}. These entities display striking analogies to matter molecules, with properties such as formation dynamics, intrinsic vibrational modes, and switching behaviours. Soliton molecules are pervasive in ultrafast lasers and passive nonlinear resonators, where multiple travelling pulses can interact over extended temporal scales. 
While soliton pairs constitute the fundamental building blocks of such assemblies, higher-order molecular complexes are also possible, including macromolecules, soliton crystals \cite{PRA_Haboucha_2008,NP_Cole_2017,NPhys_Karpov_2019,NC_Lu_2021}, and even highly ordered supramolecular arrangements through engineered long-range interactions \cite{NC_He_2019}. Remarkably, although breathing solitons are fundamentally distinct from stationary ones, recent studies have demonstrated that they can also exhibit collective behaviour akin to molecular organisation \cite{NC_Lucas_2017,SA_Peng_2019,OE_Chen_2019,OE_Wang_2019, OL_Wang_2020,APR_Zhou_2022,OL_Du_2022,LP_Lu_2023, OLT_Fan_2024, OLT_Li_2024}. In \cite{SA_Peng_2019}, using real-time temporal and spectral measurements of a normal-dispersion mode-locked fibre laser, we reported the first experimental observation of breathing soliton-pair (``diatomic'') molecules in lasers. This study was expanded in \cite{LPR_Peng_2021}, where we demonstrated a variety of breather molecular complexes in an anomalous-dispersion cavity by tuning the intra-cavity loss at fixed pump strength. These included multi-breather (tetratomic) molecules, as well as hybrid assemblies formed by the binding of two diatomic molecules or of a diatomic molecule with a single breather.
A key finding was that the inter-molecular temporal separations in these breather complexes exceeded those in stationary soliton molecules \cite{NC_Wang_2019} by more than an order of magnitude. This observation is consistent with the presence of long-range interactions \cite{PRL_Turaev_2012} mediated by slowly decaying dispersive waves radiated in the anomalous-dispersion regime \cite{OL_Du_2019}. Additionally, we observed rich non-equilibrium dynamics within these complexes, including breather collisions and the annihilation of individual breathers.

In this section, we demonstrate the automatic generation of breather molecular complexes with a controllable number of constituent breathers within a laser cavity, enabled by the use of GAs \cite{LPR_Wu_2022}. Furthermore, we extend the study of synchronisation phenomena to include multi-breather complexes \cite{PRL_Wu_2023}, uncovering new pathways for manipulating their collective dynamics.

\subsection{Intelligent dynamics generation}

\begin{figure}
\centering
{\resizebox*{12.cm}{!}{\includegraphics{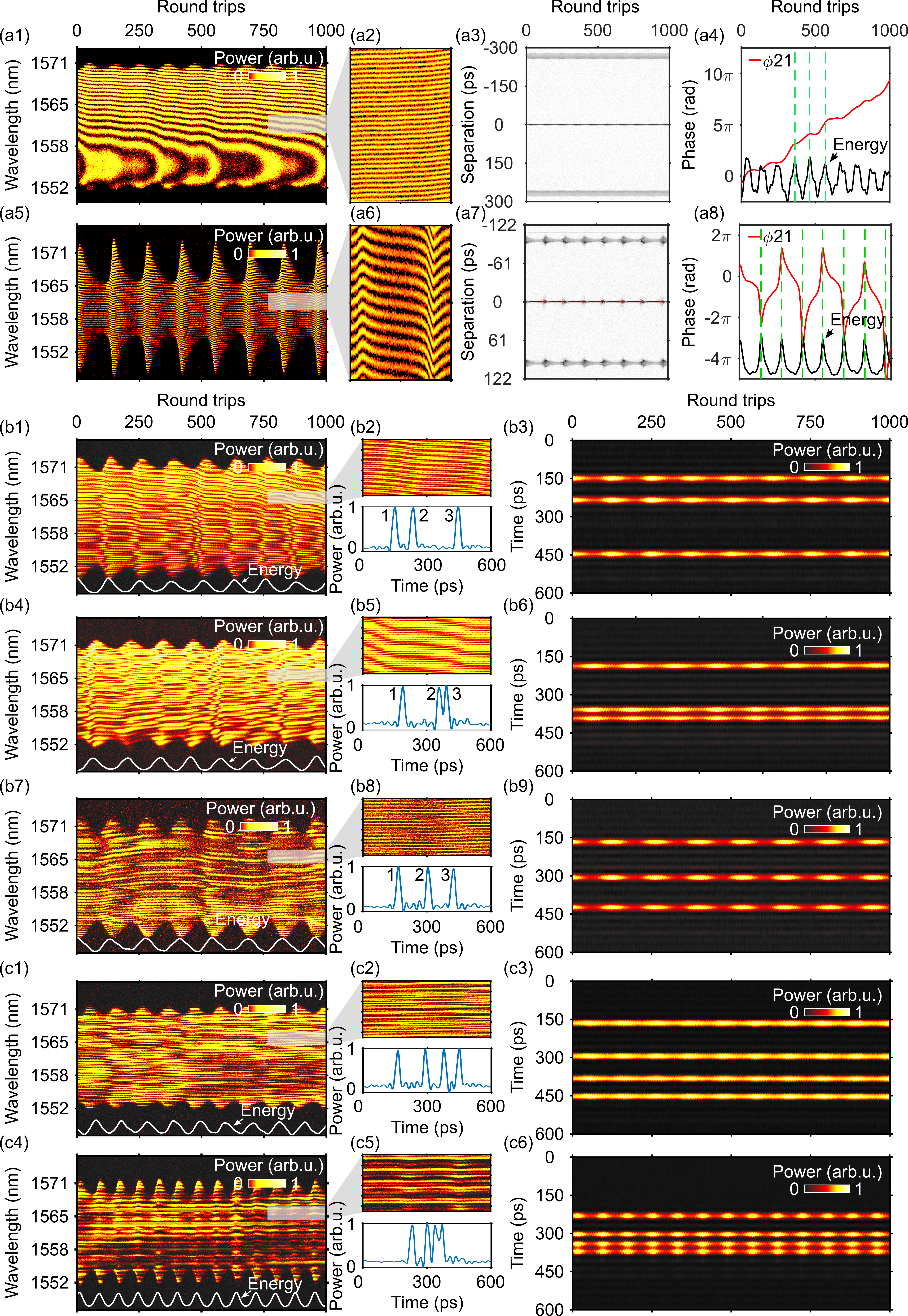}}}
\caption{{\bf Genetic-algorithm–optimised breather molecular complexes containing two, three, and four breathers. (a) Diatomic molecules.} Two archetypes are shown: an {\itshape increasing-phase} pair (top row) and an {\itshape oscillating-phase} pair (bottom row).
(a1,$\,$a5) DFT recordings of single-shot spectra over successive cavity round-trips; the dense Moir\'e fringe pattern in (a1) indicates a large intramolecular pulse separation. (a2,$\,$a6) Magnified views of the DFT data. (a3,$\,$a7) First-order single-shot autocorrelation traces versus round-trip number. (a4,$\,$a8) Evolution of the relative phase between the two breathers (red) and of the total molecule energy (black).
{\bf (b) Breather triplets.} Results for a $(2+1)$ complex, a $(1+2)$ complex, and a triatomic molecule.
(b1,$\,$b4,$\,$b7) DFT single-shot spectra; white curves trace the total intracavity energy. (b2,$\,$b5,$\,$b8) Enlarged spectral windows together with the corresponding temporal-intensity profiles. (b3,$\,$b6,$\,$b9) Temporal-intensity evolutions referenced to the average round-trip time.
{\bf (c) $\mathbf{(1 + 3)}$ complexes.} Examples with (top) large and (bottom) small internal pulse separations. (c1,$\,$c4) DFT single-shot spectra.
(c2,$\,$c5) Close-up spectral views and associated temporal intensities.
(c3,$\,$c6) Temporal-intensity evolutions relative to the average round-trip time. Adapted from \cite{LPR_Wu_2022}.} \label{BMCs}
\end{figure}

While soliton molecules in fibre lasers can be generated simply by raising the pump power above the fundamental mode-locking threshold---with the number of solitons in each molecule scaling monotonically with the pump power \cite{PRL_Krupa_2017,NC_Wang_2019}---the excitation of breather molecules is more challenging. Moreover, in normally dispersive fibre cavities, the formation of multi-breather complexes (more than two breathers) is hindered because breathers propagating in the normal-dispersion regime do not emit dispersive waves \cite{SA_Peng_2019}.
To circumvent this limitation, we recently introduced a GA strategy that optimises the highly dynamic breathing behaviour of ultrafast lasers by exploiting distinctive features in the RF spectrum of the breather output (see Section \ref{sec:GA} and Ref.\ \cite{LPR_Wu_2022}). Using this approach, we accessed a broad family of breather molecular complexes exhibiting diverse internal dynamics. Representative results are collected in Fig.\ \ref{BMCs}. In particular, Fig.\ \ref{BMCs}(a) displays two examples of breather-pair molecules, showing the round-trip evolution of their DFT spectra, first-order single-shot optical autocorrelation traces obtained via Fourier transformation of those spectra, and the relative phases between the constituent breathers extracted from the autocorrelation data \cite{NC_Wang_2019}. Figure \ref{BMCs}(a1) displays a dense fringe pattern in the single-shot spectra, exhibiting a pronounced Moir\'e effect. The very small fringe spacing corresponds to a large intramolecular pulse separation of 268$\,$ps, as confirmed by the autocorrelation trace in Fig.\ \ref{BMCs}(a3). The relative phase, $\phi_{2,1}$, between the trailing and leading breathers evolves almost linearly with the number of round-trips [red curve, Fig.\ \ref{BMCs}(a4)]. Because the phase-evolution slope is proportional to the intensity difference between the two bound pulses \cite{PRL_Krupa_2017,Science_Herink_2017}, this linear trend implies an essentially constant intensity imbalance, with the trailing breather remaining the more intense of the pair throughout the evolution.
By contrast, the molecule in Figs.\ \ref{BMCs}(a5--a8) exhibits a markedly broader breathing of the optical spectrum, an intramolecular pulse separation reduced by nearly a factor of three, and a strongly oscillatory relative-phase dynamics. The phase modulation in Fig.\ \ref{BMCs}(a8) signals continuous energy exchange between the two breathers: the pulses attain equal intensity and the total intracavity energy peaks at the round-trip indices where the phase-evolution curve reaches its extrema.

Extending the discussion beyond di-breather molecules, Fig.\ \ref{BMCs}(b) summarises GA-optimised solutions for three archetypal bound-breather triplets: a $(2 + 1)$ complex, in which a di-breather molecule precedes a single breather; a $(1+2)$ complex, where the single breather leads the di-breather; and a triatomic molecule comprising three nearly equidistant breathers. The DFT-based single-shot spectra and the corresponding spatio-temporal intensity maps differ markedly among these cases, and these contrasts govern their internal phase and energy dynamics. As shown in \cite{LPR_Wu_2022}, in the $(2 + 1)$ and $(1+2)$ complexes, the relative phases evolve almost linearly with round-trip number: the central pulse is the weakest, while the trailing pulse is strongest in the $(2+1)$ configuration; the opposite hierarchy holds for the $(1+2)$ configuration, where the leading pulse dominates. By contrast, the triatomic molecule exhibits an oscillatory phase evolution. Figure \ref{BMCs}(c) shows the dynamics of two representative $(1+3)$ breather complexes. The spatio-temporal intensity evolutions in panels (c3) and (c6) reveal markedly different pulse separations within the two complexes, highlighting the diversity of their structural configurations. As with the previously discussed cases, more targeted measurements allow us to resolve the internal motion of these complexes in detail and to discern the distinct dynamical behaviours that emerge \cite{LPR_Wu_2022}. These behaviours contrast significantly with those typically observed in stationary soliton molecular complexes, underscoring the richer and more complex internal dynamics accessible in breather-based systems.

\subsection{Synchronisation, desynchronisation and intermediate regime}
\label{sec:SynchDesynch}

Building on the investigation of single-breather synchronisation discussed in Section \ref{sec:Synchronisation}, in \cite{PRL_Wu_2023} we demonstrated, for the first time, the occurrence of subharmonic synchronisation and desynchronisation of multi-breather molecule-like bound states within a laser cavity. As noted in Section \ref{sec:Synchronisation}, we additionally identified an intermediate regime---modulated subharmonic breathing---that emerges between the synchronised and desynchronised phases. Our results underscore the inherent robustness of the phase transition, demonstrating that it persists independently of the number of constituent breathers forming the soliton structure. This finding further supports the universality of synchronisation and desynchronisation phenomena in nonlinear systems and opens new avenues for investigating the dynamics of systems involving three or more interacting frequencies \cite{PRL_Grebogi_1983,PRL_Cumming_1988}.

\begin{figure}
\centering
{\resizebox*{14.5cm}{!}{\includegraphics{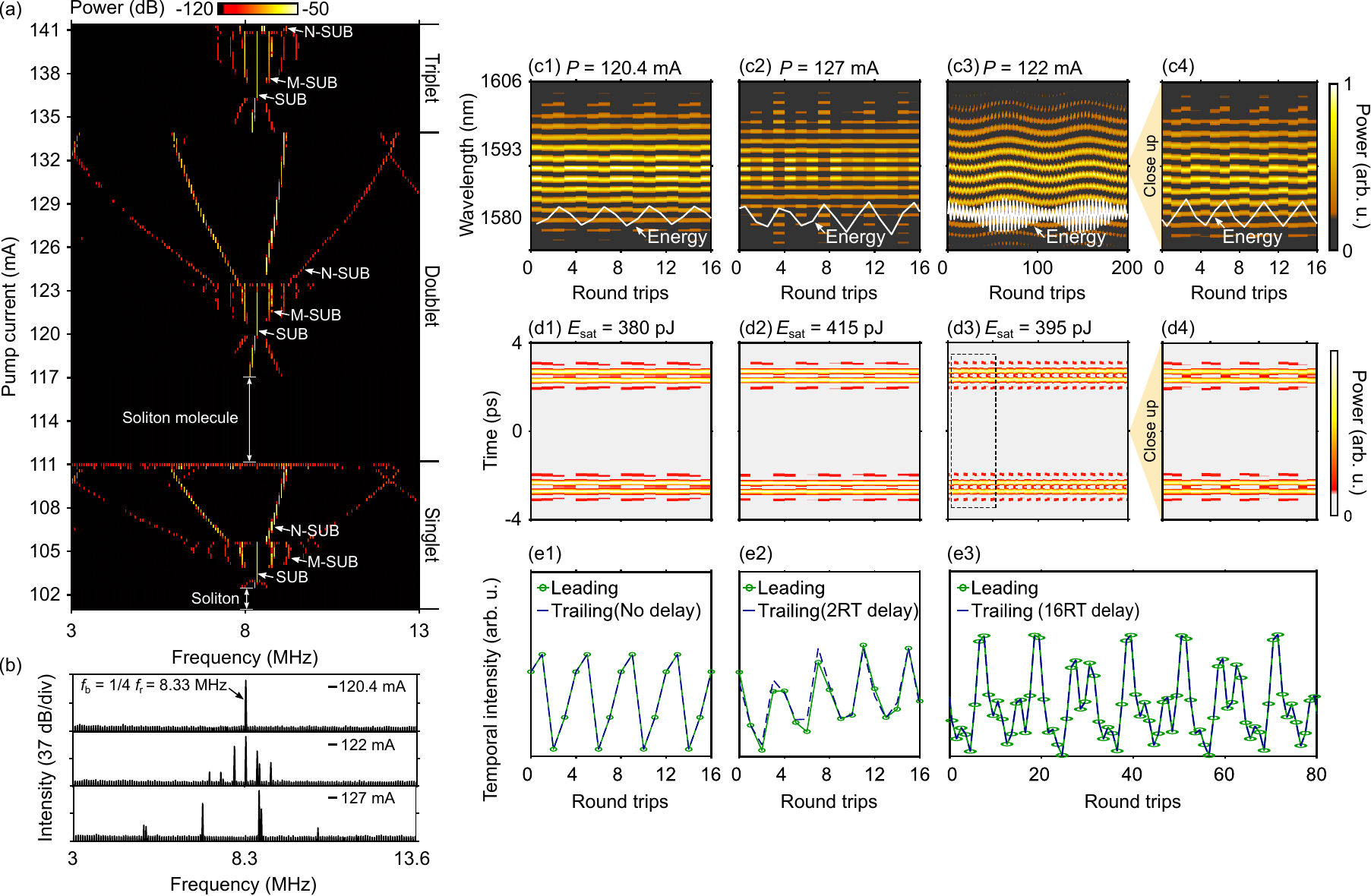}}}
\caption{{\bf Synchronisation dynamics and regimes of breathing-soliton structures}.
(a) Measured RF spectra of the laser output as a function of pump current, illustrating a sequence of dynamical transitions from subharmonic (SUB) to modulated subharmonic (M-SUB) and finally to non-subharmonic (N-SUB) states. These transitions are observed for single breathers as well as for diatomic and triatomic breather molecules.
(b1--b3) DFT recordings of single-shot optical spectra over consecutive cavity roundtrips, corresponding to SUB, N-SUB, and M-SUB regimes in diatomic breather molecules, respectively. The overlaid white traces depict the pulse energy evolution.
(b4) Magnified view of (b3), highlighting short-period breathing dynamics.
(c) Representative RF spectra for the three breathing regimes.
(d1--d3) Simulated temporal intensity evolutions over successive roundtrips for SUB, N-SUB, and M-SUB diatomic breather molecules. (d4) magnified view of (d3).
(e) Evolution of peak intensities for the leading and trailing pulses in the three breather molecule regimes. In the N-SUB and M-SUB states, the trailing pulse has been delayed by a fixed number of roundtrips to demonstrate lag synchronisation between the two breathers.   Adapted from \cite{PRL_Wu_2023}.} \label{SynchDesynch}
\end{figure}

Figure \ref{SynchDesynch}(a) presents a typical evolution of the RF spectrum of the laser emission as a function of pump current. At low currents (up to 102$\,$mA), the laser emits a single soliton pulse per cavity round trip, marked by a single frequency component at the cavity repetition frequency $f_{\mathrm{r}}=33.39\,$MHz
(not visible in Fig.\ \ref{SynchDesynch}(a), where a reduced  frequency span is used for clarity). Increasing the pump current induces a Hopf bifurcation, leading to the generation of a breathing soliton with a pulsation period of four round trips, as evidenced by the emergence of a narrow subharmonic peak at $f_{\mathrm{b}}=f_{\mathrm{r}}/4$, indicating frequency locking \cite{NC_Wu_2022}. Further pump increase drives the laser into the so-called {\itshape modulated subharmonic} regime \cite{PRL_Wu_2023}, distinguished by the appearance of additional, symmetrically spaced spectral lines around $f_{\mathrm{b}}$. The spacing between these lines corresponds to a longer pulsation period in the time domain, forming a characteristic `modulated subharmonic' RF structure. Beyond 106$\,$mA, the frequency locking is disrupted: the modulated sidebands vanish and $f_{\mathrm{b}}$ begins to drift continuously with pump current, becoming non-commensurate with $f_{\mathrm{r}}$. 
At currents exceeding 111$\,$mA, the pulsating behaviour gives way to brief chaotic dynamics, followed by the formation of diatomic stationary-soliton molecules up to approximately 117$\,$mA. Subsequent increases introduce breathing-soliton molecules, with alternating subharmonic, modulated subharmonic, and non-harmonic regimes observed between 117--133$\,$mA. Beyond 133$\,$mA, the system transitions to triatomic breather molecules, repeating the same RF spectral evolution. Figure \ref{SynchDesynch}(a) also reveals an important trend: as the number of elementary constituents increases, the subharmonic breather structures exhibit enhanced robustness against variations in pump strength.  

Complementary spatio-spectral measurements of the laser dynamics across the three phases are summarised in Fig.\ \ref{SynchDesynch}(b), using the diatomic breather molecule as a representative example. Qualitatively similar behaviour was observed for the triatomic breather molecule regime \cite{PRL_Wu_2023}. Panels (i--iii) show roundtrip-resolved optical spectra acquired via the time-stretch technique.
The subharmonic and non-subharmonic regimes exhibit strictly periodic and degraded periodic variations, respectively, in both the optical spectrum and pulse energy across successive roundtrips [cf.\ Figs.\ \ref{FrequencyLocking}(a2,\,b2)]. In both cases, the period of spectral fringes remains nearly constant, indicating a stable intra-molecular pulse separation. By contrast, the modulated subharmonic regime features two distinct periodicities: a short period of 4 roundtrips and a long modulation period of approximately 88 roundtrips. The corresponding RF spectra around $f_{\mathrm{r}}/4$ [Fig.\ \ref{SynchDesynch}(c)] further highlight the distinctions among the three regimes: the subharmonic state exhibits a single, extremely narrow frequency component located exactly at 
$f_{\mathrm{b}}=f_{\mathrm{r}}/4$; the modulated subharmonic state displays a symmetric set of narrow sidebands around $f_{\mathrm{b}}=f_{\mathrm{r}}/4$; 
while the non-subharmonic regime presents broadened spectral lines, consistent with frequency-unlocked laser operation \cite{NC_Wu_2022}.

The RF spectral intensity as a function of the gain saturation energy, obtained from simulations of the laser model (cf.\ Section \ref{LumpedModel}), showed excellent agreement with experimental observations \cite{PRL_Wu_2023}. Crucially, the simulations---providing full access to the underlying temporal dynamics---unveiled that in the unsynchronised and modulated subharmonic regimes, the constituent breathers within a soliton molecule are mutually synchronised with a constant time delay, exhibiting {\itshape lag synchronisation} \cite{Pikovsky_2001,PRL_Rosemblum_1997,PRE_Taherion_1999}. This behaviour is illustrated in Figs.\ \ref{SynchDesynch}(d,\,e), where the roundtrip evolution of the temporal intensity profiles and the peak intensities of the leading and trailing pulses reveal a consistent delay between the two breathers [panels (d2-3)]. When this delay is numerically compensated for, the evolution of the pulses becomes fully synchronous [panels (e2-3)]. Similar lag synchronisation behaviour was also observed in triatomic breather molecules. Moreover, the temporal profiles [Fig.\ \ref{SynchDesynch}(d)] highlight a recurring structural feature across the three breather molecule states: each breather is composed of multiple sub-pulses, indicative of higher-order soliton-like evolution within the anomalous dispersion segment of the laser cavity.

By adjusting the laser's polarisation state in the experiment (equivalent to varying  intracavity loss in the model), we also identified direct transitions between synchronised and desynchronised states, consistent with saddle-node bifurcations \cite{Pikovsky_2001}. These dynamical transitions resonate with broader analogies in the field of dissipative soliton physics, where nonlinear fibre lasers have been shown to exhibit behaviours reminiscent of states of matter---such as soliton molecules, crystals, rains, and gases \cite{APB_Amrani_2010}. In this context, the synchronisation–desynchronisation dynamics of breathing solitons and their bound states can be qualitatively linked to commensurate–incommensurate phase transitions \cite{RPP_Bak_1982}, a class of phenomena well-known in condensed matter physics and other complex systems.

\section{Transition to chaos and the modulated subharmonic route}
\label{sec:Chaos}
Theoretical studies have shown that solitons can exhibit chaotic behaviour in perturbed systems \cite{PLA_Eilbeck_1981, PRL_Nozaki_1983, PRL_Blow_1984, PRL_SotoCrespo_2005}. While experimental observations have so far been largely restricted to spin-wave systems \cite{PRL_Wang_2011, PRL_Ustinov_2011}, the possibility of chaos driven by optical solitons remains a subject of considerable interest. Chaotic solitons offer a natural extension of laser chaos \cite{PLA_Haken_1975} into the framework of the generalised NLSE, and this extension is significant for two main reasons. First, the generation of chaos within the Maxwell–Bloch formalism generally requires external signal injection \cite{PRL_Yamada_1980} or impractically high pump powers, whereas soliton chaos can arise spontaneously in free-running mode-locked lasers, as predicted in \cite{PRL_SotoCrespo_2005}. Second, a wide range of physical systems are known to follow three well-established universal routes to chaos: the Ruelle–Takens scenario via quasi-periodicity \cite{Ruelle_1971}, the Feigenbaum scenario via period-doubling \cite{JSP_Feigenbaum_1978}, and the Pomeau–Manneville scenario via intermittency \cite{CMP_Pomeau_1980}. These classical routes are typically described by Lorenz-type systems or one-dimensional maps. In contrast, the GNLSE is fundamentally different and may enable novel pathways for the transition from regular to chaotic dynamics. Identifying such mechanisms in real systems governed by the NLSE or its extensions is therefore of fundamental importance in the broader context of nonlinear science.

Despite the significance of soliton chaos and early theoretical predictions in mode-locked lasers \cite{PRL_SotoCrespo_2005}, experimental studies of optical chaotic solitons have remained limited—primarily due to ultrafast dynamics that exceed the temporal resolution of conventional electronic detection systems. Consequently, earlier experiments were unable to unambiguously confirm the presence of soliton chaos \cite{CIJNS_Zhao_2006}, a challenge similarly encountered in Kerr resonator systems \cite{PRL_Nielsen_2021}. However, a recent study \cite{CSF_Zhang_2024} has provided the first systematic experimental evidence of soliton-to-chaos transitions in a mode-locked laser, following a route marked by cascaded short- and long-period pulsations, as revealed through RF spectral analysis and Lyapunov exponent evaluation. More recently, this same scenario of cascaded pulsations has also been observed in the internal dynamics of soliton-pair molecules \cite{OE_Lu_2024}.

In \cite{PRL_Kang_2024}, we employed real-time measurements to reveal a new pathway from solitons to chaos, wherein the transition occurs via the modulated subharmonic state of a breather laser described in the previous section. This route to chaos---referred to as the {\itshape modulated subharmonic route}---was unambiguously identified in both experiments and numerical simulations, and its universality was demonstrated across two distinct laser architectures: figure-eight \cite{OL_Duling_1991} and ring-cavity configurations.

\begin{figure}
\centering
{\resizebox*{14.2cm}{!}{\includegraphics{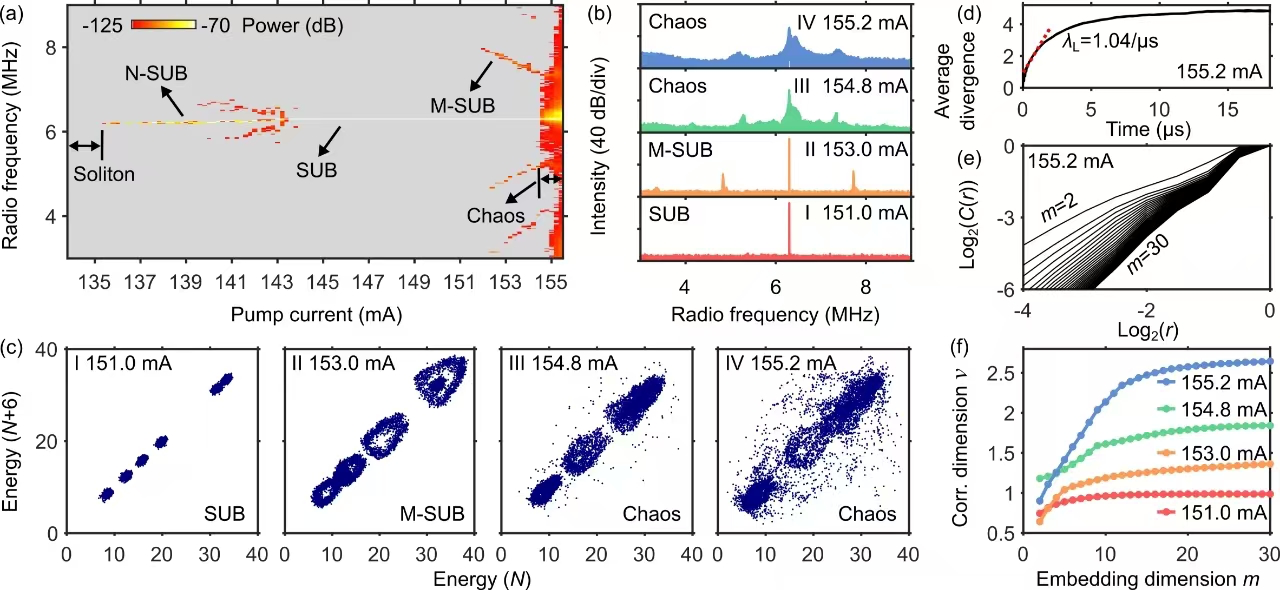}}}
\caption{{\bf Experimental observation of modulated subharmonic route to chaos}. (a) RF spectrum of the laser output as a function of pump current, illustrating successive dynamical transitions from stationary solitons to non-subharmonic (N-SUB), subharmonic (SUB), modulated subharmonic (M-SUB) breathing solitons, and eventually to chaos. The subharmonic breather state is characterised by a rational winding number of $f_{\mathrm{b}}/f_{\mathrm{r}}=1/6$. (b) Representative spectra corresponding to the SUB (151$\,$mA), M-SUB  (153$\,$mA), and chaotic (154.8 and 155.2$\,$mA) states. (c) Corresponding Poincar\'e sections showing the phase portraits of pulse energy at cavity roundtrip $N+6$ versus $N$. (d) Maximum Lyapunov exponent analysis for the chaotic state in (b),\,IV. The average divergence of nearby trajectories is fitted with $e^{\lambda_{\mathrm{L}}t}$, yielding $\lambda_{\mathrm{L}}=1.04\,\mu\mathrm{s}^{-1}$. (e) Grassberger–Procaccia analysis for the same chaotic state, presenting the correlation integral $C(r)$ versus sphere radius $r$
for different embedding dimensions $m$. The slope at small 
$r$ gives an estimate of the correlation dimension. (f) Correlation dimension as a function of embedding dimension for the four dynamical states shown in (b).  Adapted from \cite{PRL_Kang_2024}.} \label{ModulatedSubharmonicRoute}
\end{figure}

Figure \ref{ModulatedSubharmonicRoute} summarises the experimental results obtained using a figure-eight laser setup. As shown in panels (a) and (b), the characteristic RF sidebands of the modulated subharmonic state  drift  with increasing pump power and have a noisy structure. These unstable sidebands \cite{PRL_Kang_2024} ultimately give rise to chaos, as indicated by the emergence of a significantly broadened RF spectrum. To validate the chaotic behaviour, we computed Poincar\'e maps, Lyapunov exponents, and correlation dimensions of the reconstructed phase space \cite{Ott_2002,Hilborn_2000}. Figure \ref{ModulatedSubharmonicRoute}(c) presents the sequence of Poincar\'e sections corresponding to the laser operating states shown in Fig.\ \ref{ModulatedSubharmonicRoute}(b). In the subharmonic breather state (panel I), the phase portrait exhibits six fixed points, which evolve into open loops in the modulated subharmonic state (panel II). These loops become connected by scattered points (panel III), indicating an expansion of the phase space and the onset of chaos. In the fully chaotic regime (panel IV), this structure is further broadened. Notably, the Poincar\'e section in panel III retains periodic components---evidenced by open loops---coexisting with a certain amount of chaotic motion, a hallmark of the so-called {\itshape chaotic resonance} predicted in \cite{PRL_SotoCrespo_2005}. This mixed state is also reflected in the RF spectrum [Fig.\ \ref{ModulatedSubharmonicRoute}(b), III], where a dominant peak indicates periodicity, while the surrounding broadband noise signifies chaos. It is also worth noting that the phase portraits  were reconstructed using pulse energies derived from the single-shot optical spectra measured via DFT. In contrast, attempts to extract similar phase-space information from direct temporal intensity measurements---without time-stretch---failed to resolve the attractor structure due to the limited temporal resolution of the photodetector, which could not capture the soliton duration ($\sim 665\,$fs) \cite{PRL_Kang_2024}. 

The Lyapunov characteristic exponent ($\lambda_{\mathrm{L}}$) quantifies the rate at which nearby trajectories diverge in phase space; a positive value is a signature of chaos \cite{Ott_2002,Hilborn_2000}. Figure \ref{ModulatedSubharmonicRoute}(d) shows the exponential divergence of trajectories for the fully chaotic regime, with $\lambda_{\mathrm{L}}=1.04\,\mu\mathrm{s}^{-1}$. We also measured a positive $\lambda_{\mathrm{L}}=0.399\,\mu\mathrm{s}^{-1}$ for the chaotic resonance state, a negative value for the subharmonic state, and a near-zero but still slightly positive $\lambda_{\mathrm{L}}=0.045\,\mu\mathrm{s}^{-1}$ for the modulated subharmonic state, due to measurement noise \cite{PRL_Kang_2024}. To further characterise the system, we estimated the correlation dimension ($\nu$) using the Grassberger–Procaccia algorithm \cite{PRL_Grassberger_1983}. Figure \ref{ModulatedSubharmonicRoute}(e) displays the correlation integral $C(r)$ as a function of the sphere radius $r$
for varying embedding dimensions $m$, showing clear saturation of $\nu$ for the chaotic state [Fig.\ \ref{ModulatedSubharmonicRoute}(c), panel IV], indicative of deterministic chaos. This trend is summarised in Fig.\ \ref{ModulatedSubharmonicRoute}(f), where all four states exhibit saturated $\nu$ values that reflect their increasing dynamical complexity---from periodic to chaotic. 

By varying the laser's polarisation state (intracavity loss), we found that solitons in our laser can also transition to chaos via the subharmonic route---through a subharmonic breather state---previously reported in other physical systems \cite{PRL_Lauterborn_1981,PRL_DeAguiar_1986}, and the quasi-periodicity route---through a non-subharmonic breather state---echoing recent observations in magnetic films \cite{PRL_Ustinov_2011}. As a final remark, we emphasise the critical role of the polarisation controller’s settings in enabling soliton chaos; merely increasing the pump power from the soliton regime does not typically induce chaotic behaviour \cite{PRL_Kang_2024}.

\section{Conclusions and outlook}
We have presented an overview of our recent research demonstrating ultrafast fibre lasers operating in the breathing-soliton regime as a powerful and versatile platform for investigating complex synchronisation phenomena and chaotic dynamics relevant to a wide range of physical systems---all within a single nonlinear oscillator, without the need for coupled systems or external forcing. In parallel, we have introduced intelligent control of breather dynamics using GAs, enabling systematic exploration and optimisation of dynamic states.

Together, these findings advance our fundamental understanding of nonlinear dynamics and provide a novel experimental framework for probing and controlling complex systems. Notably, we report the first experimental observation of unconventional synchronisation structures that had remained unconfirmed in physical systems since the prediction of leaf-like patterns in the circle map model over 25 years ago \cite{PR_Feudel_1997}. We also uncover a modulated subharmonic regime that bridges synchrony and desynchrony, as well as a route to chaos via modulation of subharmonic states---both of which have not been previously observed in physical systems.

The synchronisation and desynchronisation behaviours observed in breather structures are qualitatively linked to commensurate–incommensurate phase transitions \cite{RPP_Bak_1982}, a class of phenomena well known in condensed matter physics. While single-breather oscillations offer a minimal system for studying two-frequency interactions, multi-breather states introduce additional degrees of freedom through their constituent breathing frequencies---paving the way for the investigation of systems with three or more interacting frequencies \cite{PRL_Grebogi_1983, PRL_Cumming_1988,PRE_Cartwright_1999}. In parallel, since nonlinear interactions among three frequencies are known to give rise to low-dimensional chaos \cite{PRL_Kang_2024, PRL_Grebogi_1983,PRL_Cumming_1988}, our results suggest that studying coupled breathing-soliton oscillators and/or multi-breather oscillators could open avenues for exploring high-dimensional chaos (hyperchaos) \cite{PLA_Rossler_1979,PRL_Fischer_1994} and synchronisation of chaos \cite{PRL_Pecora_1990,PR_Boccaletti_2002}.

In addition to the ultrafast breathing-soliton lasers discussed, recent studies have extended the breathing-soliton phenomenon to spatiotemporal mode-locked (STML) multimode fibre lasers \cite{Science_Wright_2017, NP_Wright_2020, LSA_Cao_2023}. These systems naturally exhibit complex, high-dimensional nonlinear dynamics, including periodically tunable multimode soliton pulsations \cite{OE_He_2024}, spatiotemporal soliton and soliton-molecule behaviours \cite{NC_Guo_2021, NC_Guo_2023}, and spatiotemporal period-doubling bifurcations \cite{ACS_Phot_Xiao_2022}. The presence of multiple coupled spatial and temporal modes introduces additional degrees of freedom, possibly enabling interactions among several breathing frequencies and thereby offering a fertile platform for exploring routes to hyperchaos, synchronisation and desynchronisation of complex oscillatory states, and multi-frequency commensurate-incommensurate transitions. STML fibre lasers thus offer a complementary and more richly structured setting for extending our studies of breather-mediated nonlinear dynamics. Beyond fibre systems, solid-state lasers have also been shown to support breathing vortex solitons \cite{LPR_Liu_2025}, further underscoring the generality of breathing-soliton phenomena across platforms.

Very recently, we have developed a unified model for breather fibre lasers that incorporates the spatiotemporal dynamics of the gain medium. This model reveals the distinct formation mechanisms of breathing solitons under net-normal and near-zero net cavity dispersion, which explain the markedly different experimental behaviours observed in each regime. Specifically, while a combination of Q-switching and soliton shaping is responsible for the formation of breathing solitons under net-normal dispersion, Kerr and dispersion effects dominate the generation of breathing solitons under near-zero dispersion. As partially discussed in this work, these differences manifest in the pump-power range relative to stationary mode-locking, oscillation period, spectral characteristics, and synchronisation capabilities. These insights advance our understanding of the physics governing breather dynamics and will be presented in detail in a forthcoming publication \cite{Submitted_Wu_2025}.

From an applied perspective, our study holds important implications for mastering complex laser behaviour, a key requirement for the development of high-performance, turn-key laser sources. Additionally, frequency-locked breather lasers can generate dense RF combs with sub-MHz line spacing---surpassing cavity length limitations---making them highly attractive for high-resolution spectroscopy. Moreover, Arnold tongues provide a robust mechanism for controlling synchronisation dynamics. Understanding the conditions that lead to the formation of holes within the synchronisation regions is crucial, as this knowledge enables their suppression and ensures stable and reliable system operation. Concurrently, chaotic solitons offer new capabilities for chaos-based technologies such as parallel optical ranging, by leveraging their broad spectral bandwidth, in contrast to microcomb-based systems relying on modulation instability \cite{NP_Chen_2023,NP_Lukashchuk_2023}.

The development of smart, self-optimising ultrafast fibre oscillators is gaining increasing importance \cite{Photonix_Jiang_2022, NM_Ma_2023}, as many emerging applications demand lasers with precisely tailored temporal and spectral characteristics. Traditional trial-and-error approaches to laser design and optimisation are time-consuming, often irreproducible, and poorly suited for real-time control. Meanwhile, systematic numerical propagation modelling remains computationally expensive, limiting exploration of broader subspaces of useful ultrafast dynamics. Within the broader context of smart ultrafast photonics, artificial intelligence–driven control---particularly through ML approaches---offers a promising path forward. Neural networks \cite{OLT_Boscolo_2020, NMI_Salmela_2021, LPR_Pu_2023}, and especially physics-informed learning frameworks \cite{JCP_Raissi_2019}, which require less training data and offer greater robustness than purely data-driven models, provide computationally efficient tools for solving both forward and inverse problems in nonlinear systems. These approaches hold strong potential for discovering novel phenomena and deepening our understanding of underlying physical mechanisms. In parallel, advanced signal processing techniques are emerging as powerful tools for analysing and characterising the complex radiation dynamics of fibre lasers. One such technique is the inverse scattering transform based on the Zakharov–Shabat system \cite{JETP_Zhakarov_1972}---commonly referred to in optics as the nonlinear Fourier transform---which is applicable in regimes where coherent, localised structures are embedded within dispersive backgrounds \cite{OC_Wang_2023,NP_Ryczkowski_2018,NP_Sugavanam_2019}. Additionally, methods such as dynamic mode decomposition and the sparse identification of nonlinear dynamical systems (SINDy) algorithm have shown promise in analysing the internal dynamics of soliton molecules \cite{OL_Sheveleva_2023, CSF_Sheveleva_2025} and may be extended to characterise a wide range of periodic and multi-scale nonlinear interactions. 

We anticipate that our work will stimulate further research efforts in these directions, both within our group and across the broader ultrafast photonics and nonlinear dynamics communities.

\section*{Disclosure statement}
The authors declare that they have no competing interests. 

\section*{Funding}
We acknowledge support from the Innovation Program for Quantum Science and
Technology (2023ZD0301000); the National Natural Science Fund of China (12434018, 62475073, 1243000542, 11621404, 11561121003, 11727812, 61775059, 12074122, 62405090, 62035005, 11704123); the Shanghai Natural Science Foundation (23ZR1419000); the China Postdoctoral Science Foundation (2023 M741188, 2024 T170275); the Agence Nationale de la Recherche (ANR-20-CE30-004); the Region Bourgogne-Franche-Comt\'e; the Institut Universitaire de France; and the NATO Science for Peace and Security Programme (SPS.MYP.G6137 -- MELITE).

\bibliographystyle{tfnlm}
\bibliography{interactnlmsample}

\end{document}